\documentclass[aps,prd,reprint,twocolumn,preprintnumbers,floatfix,nofootinbib]{revtex4-1}
\usepackage[utf8]{inputenc}
\pdfoutput=1
\usepackage{graphicx}
\usepackage{amsmath}
\usepackage{amsfonts}
\usepackage{amssymb}
\usepackage{xcolor,soul}
\usepackage{epstopdf}
\usepackage{xcolor}
\usepackage{float}
\usepackage{subfigure}

\usepackage{relsize}
\usepackage{graphics}
\usepackage{hyperref}
\usepackage{mathrsfs}
\usepackage{amssymb}
\usepackage{booktabs}
\usepackage[normalem]{ulem}
\usepackage{amsthm}
\usepackage{cancel}
\usepackage{fontawesome}   
\usepackage{tabularx,ragged2e} % new

\hypersetup{
     colorlinks   = true,
     citecolor    = red,
     linkcolor    = blue,
     urlcolor     = blue,
}
\usepackage{physics}  %% DK
\newcommand{\bea}{\begin{aligned}}
\newcommand{\eea}{\end{aligned}}
\def\beq{\begin{equation}}
\def\eeq{\end{equation}}
\def\beqa{\begin{eqnarray}}
\def\eeqa{\end{eqnarray}}
\def\be{\begin{equation}}
\def\ee{\end{equation}}

\def\bse{\begin{subequations}}
\def\ese{\end{subequations}}

\def\trh{$T_{\mathrm{RH}}~$}

\def\tin{t_{\rm in}}
\def\tev{t_{\rm ev}}
\def\ain{a_{\rm in}}
\def\ae{a_{\mathrm{end}}}
\def\arh{a_{\mathrm{RH}}}
\def\abh{a_{\mathrm{BH}}}

\def\trh{T_{\mathrm{RH}}}
\def\bea{\begin{eqnarray}}
\def\eea{\end{eqnarray}}
\def\rhoe{\rho_{\mathrm{end}}}

\def\aev{a_{\mathrm{ev}}}
\def\Min{M_{\mathrm{in}}}
\def\BH{\mathrm{BH}}
\def\tbhin{T_{\rm BH}^{\rm in}}
\def\tbh{T_{\rm BH}}

\newcommand{\red}{\color{red}}
\newcommand{\blue}{\color{blue}}

\newcommand{\magenta}{\color{magenta}}

\newcommand{\rbh}{\rho_{\rm BH}}
\newcommand{\rphi}{\rho_{\phi}}

\graphicspath{{./figs/}}

\usepackage{pgfplots}
\pgfplotsset{compat=1.17}

\begin{document}
%\preprint{APS/123-QED}

%\thanks{}%

%\author{ }
%\email{ }
%\author{ }
%\email{ }
%\author{ }
%\email{ }

%\affiliation{$^a$ }
%\affiliation{$^b$ }
%\affiliation{$^c$ }

\preprint{}
\preprint{}

\vspace*{1mm}

\title{Primordial black hole versus inflaton}

\author{Md Riajul Haque$^{a}$}
\email{riaj.0009@gmail.com}

\author{Essodjolo Kpatcha$^{b,c}$}
\email{kpatcha@ijclab.in2p3.fr}

\author{Debaprasad Maity$^{d}$}
\email{debu@iitg.ac.in}

\author{Yann Mambrini$^{b}$}
\email{yann.mambrini@ijclab.in2p3.fr}

\vspace{0.1cm}

 \affiliation{
${}^a$
Centre for Strings, Gravitation, and Cosmology,
Department of Physics, Indian Institute of Technology Madras, 
Chennai~600036, India
}

\affiliation{
${}^b$ Universit\'e Paris-Saclay, CNRS/IN2P3, IJCLab, 91405 Orsay, France
 }

\affiliation{${}^c$ Departamento de F\'{\i}sica Te\'{o}rica, Universidad Aut\'{o}noma de Madrid (UAM),
Campus de Cantoblanco, 28049 Madrid, Spain}

\affiliation{
${}^d$
Department of Physics, Indian Institute of Technology Guwahati, Guwahati, Assam, India
}

\begin{abstract} 

We compare the dark matter(DM) production processes and its parameters space in the background of reheating obtained from two chief systems in the early Universe: the inflaton $\phi$ and the primordial black holes (PBHs). We concentrated on the mechanism where DMs are universally produced only from the PBH decay and the generation of the standard model plasma from both inflton and PBHs. 
Whereas the distribution of Primordial Black Holes behaves like dust, the inflaton phenomenology depends strongly on its equation of state after the inflationary phase, which in turn is conditioned by the nature of the potential $V(\phi)$. Depending upon the initial mass and population of PBHs, a large range of DM mass is shown to be viable if reheating is controlled by PBHs itself. Inflaton-dominated reheating is observed to further widen such possibilities depending on the initial population of black holes and its mass as well as the coupling of the inflaton to the standard model sector.

\end{abstract}

\maketitle

\tableofcontents

\section{Introduction}

In 1974, Hawking's proposition that black holes emit
radiation was one of the most important results of the last century \cite{Hawking:1974rv,Hawking:1975vcx}. 
It unified thermodynamics, quantum field theory, and general relativity. Only primordial black holes
(PBHs), 
formed in the first instant of the Universe can emit radiation sufficiently important to have some 
observable effects.
On the other hand, the early Universe is also supposed to be populated by a homogeneous field $\phi$ 
called inflaton \cite{Olive:1989nu,Lyth:1998xn,Martin:2013tda,Martin:2013nzq,reviews}. Its decay can also fuel the radiation 
bath through a mechanism called reheating \cite{mybook}. 
The reheating can be non-perturbative in a process called preheating \cite{Dolgov:1982th,Kofman:1997yn,Garcia:2021iag,Lebedev:2023zgw} 
or perturbative \cite{Garcia:2020wiy,Garcia:2020eof,Haque:2020zco}.
In any case, at the very end of inflation, the energy density stored in the inflaton is of the order
$\rho_{\rm end}^{\frac{1}{4}}\simeq 10^{15}$ GeV.
Such energy corresponds to the horizon mass of the order of 1 gram, where PBHs could have been formed.

Such light black holes have a very short lifetime and can subsequently reheat the Universe, entering in competition with the inflaton.
Indeed, recently, it has been shown that the PBHs are capable of reheating the Universe without the need to dominate its energy budget \cite{RiajulHaque:2023cqe}. The main reason for this phenomenon lies in the inflaton equation of state after inflation, which is related to the behavior of the inflationary potential at the minima. Whereas PBHs behave like dust with a density $\rbh\propto a^{-3}$ ($a$ being the scale factor), the inflaton $\phi$ redshifts faster for inflation equation of state
$w_\phi >0$, as 
$\rphi \propto a^{-3(1+w_\phi)}$ where $w_\phi$
is defined by the equation of state $P_\phi=w_\phi \rphi$. Moreover, the PBH decay width\footnote{throughout our work, we will consider $M_P = 1/\sqrt{8\pi G} \simeq 2.435 \times 10^{18}$ GeV as the reduced Planck mass.} 
$\Gamma_{\rm BH} \sim {M_P^4}/{M_{\rm BH}^3}$ 
{\it increases} with time as their mass $M_{\rm BH}$ decreases due to evaporation whereas $\Gamma_\phi$ {\it decreases} with time as the condensate transfers its energy to the plasma. For instance, in the case of a Yukawa type coupling $y_\phi \phi \bar f f$ between the inflaton and the Standard Model (SM), $\Gamma_\phi \propto \rphi^{\frac{2 w_\phi}{1+w_\phi}}$ \cite{Garcia:2020wiy,Garcia:2020eof}. In other words, the inflaton decay is more efficient {\it at the beginning} of the reheating process, whereas the PBHs are more efficient {\it at the end} of their lifetime \cite{RiajulHaque:2023cqe}. These two characteristics, a more diluted and less efficient inflaton with time, help to understand the possibility for the PBH to control the evolution of the temperature in the early Universe. One important point is to note that if $w_\phi>0.60$, that means for a reheating scenario with a very steep
inflaton potential, the BBN bounds of the primordial gravitational wave provide a restriction on the lower limit of the reheating temperature \cite{Haque:2021dha,Chakraborty:2023ocr}.

In the meantime, since the earlier work of Fritz Zwicky,
\cite{Zwicky:1933gu}, the presence of dark matter (DM) has been confirmed at several scales but still not discovered. One of the possibilities is that the heavy mass PBHs can be a candidate of DM \cite{Belotsky:2014kca}. 
The recent limits coming from direct detection experiments and the lack of galactic or intergalactic 
signals put severe pressure on the conventional WIMP paradigm \cite{Arcadi:2017kky}. The scenario where WIMP is produced during the reheating has recently been shown to relax such pressure and allow large parameter regions that 
can be explored in the near future \cite{Haque:2023yra}.
Constrained by the direct detection in the conventional scenario, it is imperative to look for some alternative mechanism, particularly for the DM being coupled extremely weakly with the Standard Model.
One possibility is feeble interactions with the SM. 
FIMP candidates \cite{Bernal:2017kxu} can be obtained by the exchange of heavy mediators \cite{FIMPheavy,FIMPheavy1}
or gravitons \cite{gravity,gravityriaj,gravityclery,gravityriaj2}.
However, there exists the possibility that DM is produced even before the existence of the thermal plasma.
Indeed, two energy sources are present at the very end of inflation: inflaton and PBHs, because unstable, are perfect candidates to populate the dark sector. If they both can reheat the Universe, it seems natural to wonder if they can also populate the dark sector. Indeed, they have in common that the inflaton is not charged under the Standard Model, and the PBHs have only gravitational interaction; they should not distinguish the production of a thermal bath and the decay into DM.
Note that the same argument can be used to solve the problem of leptogenesis from PBHs and has been recently nicely addressed in \cite{Calabrese:2023key}. Moreover, there are effects of Hawking's evaporation of primordial black holes (PBHs) on the DM models \cite{Gondolo:2020uqv,Sandick:2021gew}. 

One should then address the issue that the presence of PBHs in an inflaton background would not overproduce the DM.
To avoid an overclosure of the Universe, we expect to have 
constraints on the PBH parameters, namely the fraction $\beta = \frac{\rbh}{\rphi}|_{formation}$ (ratio between the PBH energy density over inflaton energy density at the point of formation)
and the formation mass $\Min$ of the PBHs formed during the reheating period.
Indeed, whereas $\beta$ determined the density 
of PBHs in the Universe, $\Min$ gives the lifetime 
of these PBHs to be compared with the inflaton decay rate. For a given DM mass $m_{j}$, a larger BH mass $\Min$ ensures a smaller width $\Gamma_{\rm BH}$, which in turn can avoid overproduction. On the other hand, for a given 
$\Min$, a DM mass larger than the initial 
black hole temperature $\tbhin$ can sufficiently suppress the production rate until the time $T_{\rm BH}\gtrsim m_{j}$ also to avoid the overproduction. 

Combining these properties of the PBHs with the richness of the inflaton phenomenology opens a large window of parameter space, which 
was closed without considering the presence of PBHs. We expect the two systems of 
the world, inflaton, and PBH, to be highly intertwined when calculating the relic density associated with the reheating process.
This is exactly the issue we want to address in this paper, organized as follows.
After a brief reminder of the reheating mechanism 
in the presence of PBH, we compute the relic abundance of DM in different scenarios in section \ref{relic}, where we solve and analyze the set of Friedmann equations. In section \ref{refinement},
we look into the case of extended mass function and the limit on the DM mass if one considers the constraints from the warm dark matter. We then discuss the influence of exact greybody factors before concluding in section \ref{conclusion}.

\section{Particle productions through PBH} \label{relic}

\subsection{Generalities}

Being interested in the production of dark matter particles from the evaporation of PBH, we propose first to summarize the main results, which will be 
useful for our analysis. Even if one can find them 
in the literature, it is somewhat convenient here 
to gather the more important equations as they are 
quite dispersed, see \cite{Baldes:2020nuv,Bernal:2020kse,Cheek:2022mmy,Cheek:2022dbx} for 
instance. We particularly want to drive the attention of the reader
to the specific references \cite{Cheek:2021odj}, \cite{Masina:2020xhk} and \cite{Green:2020jor}, which contain elaborate discussions on the PBH evolution in the early Universe. 
We also want to add the reference of B. Carr himself \cite{Carr:2003bj} 
which
comes back to the historical aspects of the discovery of PBH's evolution.
Note that none of these references work 
in a classical background dominated by the inflaton 
field, so we had to adapt the results to our 
specific environment.
To compute the relic abundance while {\it at the same time} ensuring a reheating through the combined inflaton-PBH sources, one has to solve the set of Friedmann and Boltzmann equations

\bea
&&  \frac{d\rho_\phi}{da}+3(1+w_\phi)\frac{\rho_\phi}{a}=-\frac{\Gamma_\phi}{H}\,\left(1+w_\phi\right)\,\frac{\rho_\phi}{a}
\nonumber
\\
&&
\frac{d{\rho}_R}{da}+4 \frac{\rho_R}{a}=-\frac{\rho_\text{BH}}{M_\text{BH}}\,\frac{dM_\text{BH}}{da}\, 
%\theta(a-a_{\rm in})\,\theta(a_{\rm ev}-a)
+\frac{\Gamma_\phi\,\rho_\phi\,(1+w_\phi)}{a\,H}
\label{Eq:rhoR}
\\
&&
\frac{d{\rho}_\text{BH}}{da}+3\frac{\rho_{\rm BH}}{a}=\frac{\rho_{\rm BH}}{M_\text{BH}}\,\frac{dM_\text{BH}}{da}
\nonumber
\\
&&
\frac{d{n}_{\rm S}^\text{BH}}{da}+3\frac{n_{\rm S}^{\rm BH}}{a}=\Gamma_{\text{BH}\to j}\,\frac{{\rho}_\text{BH}}{M_\text{BH}}\,\frac{1}{a\,H}
\nonumber
\\
&&
\frac{dM_{\rm BH}}{da} = - \epsilon \frac{M^4_P}{M_{\rm BH}^2}\,\frac{1}{a\,H}\,,
\label{Eq:mbh}
\nonumber
\\
&&
 3H^2 M_P^2 \;=\; \rho_{\phi}+\rho_{R} + \rho_{\rm BH}  \,,
\label{Eq:hub} 
\nonumber
\eea

\noindent
where $\Gamma_{\text{BH}\to j}$ is the BH decay width associated with dark matter particles and $\epsilon=\frac{27}{4} \frac{g_*(T_{\rm BH})\, \pi}{480}$  to the $geometric-
optics$ limit \cite{Baldes:2020nuv}. $g_*(T_{\rm BH})$ is the number of degrees of 
freedom at $T_{\rm BH}$ (106.75 for the Standard Model). $\rho_R$ is the radiation energy density whereas $n_S$ is the number density of the dark species $S$ that we will suppose scalar throughout 
our study.

Solving Eq.(\ref{Eq:mbh}) gives 
\beq
M_{\rm BH}=\Min\left(1 - \Gamma_{\rm BH} (t-\tin)\right)^{\frac{1}{3}}
\label{Eq:mbht}
\eeq
where $\tin$ is the time of formation of the PBH
of initial mass $\Min$, and

\beq
\Gamma_{\rm BH}= 3 \times \frac{27}{4} \times \frac{g_* (T_{\rm BH}) \,\pi}{480}\frac{M_P^4}{\Min^3}=3 \,\epsilon \frac{M_P^4}{M_{\rm in}^3}
\label{Eq:gammabh}
\eeq
its width. $\Min$ is defined by

\beq 
\Min=\frac{4}{3}\pi\gamma H_{\rm in}^{-3}\rho_\phi(\ain)
=4\pi\gamma M_P^2 H_{\rm in}^{-1}\,,
\label{Eq:min}
\eeq
where $\gamma=w_\phi^{{3}/{2}}$ parameterizes 
the efficiency of the collapse to form PBHs \cite{Carr:1974nx}. 

\subsection{Particle production}
The production rate of any particular species from a BH depends on its intrinsic properties, namely mass and spin. For simplicity, we will consider the spin-zero Schwarzschild BH throughout.  The emission rate of a particle of species $j$ with internal degrees of freedom $g_j$ and mass $m_j$ escaping the Schwarzschild horizon of radius $R_S$ per unit of time and energy interval is expressed as,
\bea
&&
\frac{d^2 N_j}{dt dE}= \frac{27}{4}\pi R_S^2
\times
\frac{g_j}{2 \pi^2}\frac{E^2}{e^{\frac{E}{\tbh}}\pm 1}
\nonumber
\eea
with $R_S=\frac{M_{\rm BH}}{4 \pi M_P^2}$ 
and
\beq
\tbh=\frac{M_P^2}{M_{\rm BH}}\simeq 10^{13} \left(\frac{1\rm g}{\Min}\right)~{\rm GeV}\,,
\eeq
which implies
\beq
\frac{dN_j}{dt}=\frac{27}{4}\frac{g_j\zeta(3)}{16 \pi^3}\frac{M_P^2}{M_{\rm BH}(t)}\,.
\label{Eq:dnjdt}
\eeq
Depending upon the PBH masses, we then have two distinct cases. If
$m_j \lesssim T_{\rm BH}^{\rm in}$, which is the BH temperature at its formation time, we can consider that the production is effective throughout the entire lifetime of the PBH under consideration. Integrating Eq.(\ref{Eq:dnjdt}) between $\tin$ and the evaporation time $\tev=\Gamma_{\rm BH}^{-1}$, and using the relation (\ref{Eq:mbht}), we obtain,
\beq
N_j^{m_j < \tbh^{\rm in}}=\int_{\tin}^{\tev} \frac{dN_j}{dt}=
\frac{15 g_j \zeta(3)}{g_* \pi^4}\frac{\Min^2}{M_P^2}\simeq 10^8\left(\frac{\Min} {1~\rm g}\right)^2\,.
\label{Eq:ntotmjless}
\eeq
From the expression, a simple estimation suggests that a BH of mass 10 g can produce $\sim 10$ billion particles during its lifetime.
Note that this result is valid for a scalar species
$j$ and should be multiplied by $\frac{3}{4}$ for
a fermionic dark matter. The second distinct case arises, if $m_j \gtrsim \tbh^{\rm in}$, and for such case one needs to integrate Eq.(\ref{Eq:dnjdt}) between the time $t_j$ to $\tev$, where  
$t_j$ corresponds to the time when the mass of the emitted particle satisfies $m_j =\tbh$. A straightforward calculation gives
\beq
t_j = \Gamma_{\rm BH}^{-1}\left(1 - \frac{M_P^6}{m_j^3\Min^3}\right)\,,
\eeq
and therefore, the total number of emitted particles turns out to be,
\beq
N_j^{m_j > \tbh^{\rm in}}=\int_{t_j}^{\tev}\frac{d N_j}{dt}
=\frac{15 g_i \zeta(3)}{g_* \pi^4}\frac{M_P^2}{m_j^2}\simeq 10^{14}\left(\frac{10^{10}\rm GeV}{m_j}\right)^2\,.
\label{Eq:ntotmjmore}
\eeq
We should also mention that the above result should be multiplied by $\frac{3}{4}$ for a fermionic dark matter.

To this end, a noticeable difference between the two distinct cases is worth summarizing. When the mass of the emitted particle is smaller than the black hole formation temperature, the total number of emitted particle $N_j \propto M_{\rm in}^2$ solely depend on the PBH initial mass. Otherwise, the mass of the emitted particle controls $N_j \propto m_j^{-2}$ not the mass of the PBH. 
%From this one can indeed conclude that for the first case suggest that for the latter case, one signficantly lower range of mass of the DM to satisfy the current abundance. 

\subsection{Relic abundance}

To obtain the dark matter (DM) relic abundance of the species $j$
today at $T_0$, we use \cite{mybook}, 

\beq
\Omega_jh^2= 1.6\times 10^8\,\frac{g_0}{g_{\rm RH}}\frac{N_j\times n_{\rm BH}(\aev)}{\trh^3}\,\left(\frac{\aev}{\arh}\right)^3\frac{m_j}{\text{GeV}}\,,
\label{Eq:omegah2}  
\eeq
%\beq
%\Omega_jh^2= 1.6\times %10^8\,\frac{g(T_0)}{g(\trh)}\,\frac{N_j\times n_{\rm BH}(\arh)}{\trh^3}\,\frac{m_j}{\text{GeV}}\,,
%\label{Eq:omegah2}  
%\eeq
where $n_{\rm BH}={\rbh}/{M_{\rm BH}}$ is the density of PBH. $g_{\rm RH}=106.75$ and $g_0=3.91$ are the effective number of light species for entropy at the end of reheating and present-day, respectively. We took both effective numbers of degrees of freedom for entropy and radiation is the same.  Note that for $\aev \geq \arh$, the
reheating being completed by PBHs, 
one sets $\arh=\aev$ in Eq.(\ref{Eq:omegah2}).
One then needs to determine
the PBH density $n_{\rm BH}(\aev)$ as a function of the equation of state of the inflaton.
%\textcolor{magenta}{one needs to know the number density of PBH $n_{\rm BH}$
%at the evaporation point $\aev$.
%The number density of dark matter (DM) generated by PBH decay is simply given
%by the number density of PBH at the evaporation time 
%$n_{\rm BH}$ multiplied by $N_j$.} 

The PBH behaving like dust evolves as
\beq \label{Eq:bhenergyden}
\rbh(\aev)=\beta \rho_\phi(\ain)\left(\frac{\ain}{\aev}\right)^3
%\nonumber
%\\
%&&
=48 \pi^2 \gamma^2 \beta \frac{M_P^6}{\Min^2}\left(\frac{\ain}{\aev}\right)^3
\,,
\eeq
where we used Eq.(\ref{Eq:min}) to write
\beq
\rphi(\ain)=48 \pi^2 \gamma^2 \frac{M_P^6}{\Min^2}\,.
\label{Eq:rhophiin}
\eeq
%\textcolor{magenta}{The evolution between $\ain$ and $\aev$ depends on the 
%equation of state for $\rphi$} 
Considering that PBHs formed and evaporate during inflaton domination,
we obtain

\bea
&&
\left(\frac{\ain}{\aev}\right)^3
=\left(\frac{H_{\rm ev}}{H_{\rm in}}\right)^{\frac{2}{1+w_\phi}}=\left(\frac{2}{3(1+w_\phi)}
\frac{\Gamma_{\rm BH}}{H_{\rm in}}\right)^\frac{2}{1+w_\phi}
\nonumber
\\
&&
=\left(\frac{\epsilon}{2\,(1+w_\phi)\,\pi\,\gamma}\,\frac{M_P^2}{\Min^2}\right)^\frac{2}{1+w_\phi}\,,
\label{Eq:ainaev}
\eea
where we used Eqs.(\ref{Eq:gammabh}) and (\ref{Eq:min}) for the last equality.
Combining Eqs.(\ref{Eq:bhenergyden}) and (\ref{Eq:ainaev})
one obtains for $n_{\rm BH}(\aev)={\rbh(\aev)}/{M_{\rm BH}}$
\beq 
n_{\rm BH}(\aev)= 48 \pi^2
\beta \left(\frac{\gamma^{w_\phi}\,\epsilon}{2\pi\,(1+w_\phi)}\right)^\frac{2}{1+w_\phi}
M_P^3\left(\frac{M_P}{\Min}\right)^\frac{7+3w_\phi}{1+w_\phi}.
\label{Eq:nbhaev}
\eeq
To compute the relic abundance (\ref{Eq:omegah2}), one needs 
%the DM number density $n_j=N_j\times n_{\rm BH}$ at $\aev$. However, 
to know the running between $\aev$ and $\arh$ which
depends strongly on which system, PBHs or $\phi$ leads
the reheating process. It is
necessary to distinguish the two cases explicitly.
%depending if the reheating is achieved by the PBHs population known as PBH reheating, or the coupling between inflaton to the radiation field. 
Moreover, the PBH reheating can be achieved in two different regimes for $\beta$ : if $\beta$ is larger than
a critical value $\beta_c$, given by \cite{RiajulHaque:2023cqe} 

\begin{eqnarray}
%&&\beta \gtrsim 10^{-\frac{20 w}{1+w}}\,. \\
\label{Eq:betamin}
\beta_c & = & \left(\frac{\epsilon}{(1+w_\phi)2 \pi \gamma}\right)^{\frac{2w_\phi}{1+w_\phi}} \left(\frac{M_P}{\Min}\right)^{\frac{4w_\phi}{1+w_\phi}} \,,
% & \sim &  \left(\frac{\epsilon}{(1+w_\phi)2 \pi \gamma}\right)^{\frac{2w_\phi}{1+w_\phi}} \left(\frac{M_P}{2\times 10^{5} M_P}\right)^{\frac{4w_\phi}{1+w_\phi}}
\end{eqnarray}
PBHs dominate not only the reheating, 
but also the background dynamics over the inflaton, and evaporation takes place during PBH domination.
On the other hand, if
$\beta<\beta_c$, the PBH evaporates during inflaton domination 
and can complete the reheating if $w_\phi > 1/3$, and the inflaton coupling with radiation field is smaller than some critical value\footnote{Note that in our analysis, we considered the 
inflaton reheating through its fermionic decay. Other processes have been studied in \cite{Garcia:2020wiy,Garcia:2020eof} are left for future work.} \cite{RiajulHaque:2023cqe}.
We will distinguish these four scenarios in detail in our following discussion.

\subsection{PBH reheating}
\subsubsection{$\beta>\beta_c$}
For $\beta>\beta_c$, the PBHs dominate the universe before 
their decay and complete the reheating independently of the inflaton system. As a 
consequence, the evaporation time {\it is} the reheating time, and $\aev=\arh$ in Eq.(\ref{Eq:omegah2}). 
The reheating
temperature $\trh$ is given by the condition
\bea
&&
H_{\rm RH}^2=\frac{\rho_{\rm RH}}{3M_P^2}
=\frac{g_{\rm RH} \pi^2 \trh^4}{90 M_P^2}
=\frac{4 \Gamma_{\rm BH}^2}{9}
\nonumber
\\
&&
\Rightarrow \trh^3=\frac{M_P^{\frac{15}{2}}}{\Min^{\frac{9}{2}}}\left(\frac{12\,\epsilon^2}{\alpha_T}\right)^{\frac{3}{4}}\,,
\label{Eq:trh1}
\eea
where we used Eq.(\ref{Eq:gammabh}) and the fact that PBH decay happens in a dust-dominated Universe,
$H(t_{\rm ev})=\frac{2}{3\,t_{\rm ev}}$ and $\alpha_T=\frac{\pi^2}{30}\,g_{\rm RH}$. However, in this scenario, PBHs are formed during inflaton domination, and the decay processes occur in PBH domination. Therefore, the relevant PBH mass evolution equation is
\beq
M_{\rm BH}^3(a) \simeq \Min^3-
\frac{2 \sqrt{3} \epsilon M_P^5}{\sqrt{\rho_\phi(\abh)}}\left(\frac{a}{\abh}\right)^\frac{3}{2}\,,
\eeq
where $a_{\rm BH}$ is the time
when the PBH begins to dominate the energy budget, $\rho_\phi(\abh)=\rho_{\rm BH}(\abh)$. We assumed $M_{\rm BH}(\abh)\simeq \Min$ and $a \gg \abh$.
To obtain the scale factor associated with the evaporation point, one needs to solve

%%%%%%%%%%%%%%%%%%%%%
\bea 
M(\aev)=0~~\Rightarrow ~~
&&
\frac{\aev}{\abh}=
\frac{\Min^2\rho_\phi^{\frac{1}{3}}(\abh)}{(2 \sqrt{3}\epsilon M_P^5)^\frac{2}{3}}
\label{Eq:aevpbh}
\\
&&=\frac{\Min^2\rho_\phi^{\frac{1}{3}}(\ain)}{(2 \sqrt{3}\epsilon M_P^5)^\frac{2}{3}}
\left(\frac{\ain}{\abh}\right)^{(1+w_\phi)}\,.
\nonumber
\eea
Using
\bea
\rho_\phi(\abh)&=&\rho_{\rm BH}(\abh)=
\rho_\phi(\ain)
\left(\frac{\ain}{\abh}\right)^{3(1+w_\phi)}
\nonumber
\\
&=&
\frac{1}{\beta}\rho_{\rm BH}(\ain)
\left(\frac{\ain}{\abh}\right)^{3(1+w_\phi)}
\,,
\eea
we deduce

\beq
\frac{\ain}{\abh}=\beta^{\frac{1}{3 w_\phi}}
\,,
\label{Eq:ainoverabh}
\eeq
and then, combining 
Eqs.(\ref{Eq:rhophiin}), (\ref{Eq:aevpbh}) and (\ref{Eq:ainoverabh})
we obtain

\beq \label{Eq:aevain}
\frac{\ain}{\aev}=\frac{\ain}{\abh}\frac{\abh}{\aev}=\frac{1}{\beta^\frac{1}{3}}\left(\frac{\epsilon}{2\,\pi\,\gamma}\right)^{\frac{2}{3}}\left(\frac{M_{ p}}{\Min}\right)^{\frac{4}{3}}\,.
\eeq

\noindent
Plugging Eq.(\ref{Eq:aevain}) into (\ref{Eq:bhenergyden}), we obtain for the PBH number density at the evaporation point
\beq \label{Eq:nbhev}
n_{\rm BH} (\aev)=12\,\epsilon^2\frac{M_P^{10}}{\Min^7}\,.
\eeq

\begin{figure}
    \centering
    \includegraphics[scale=0.47]{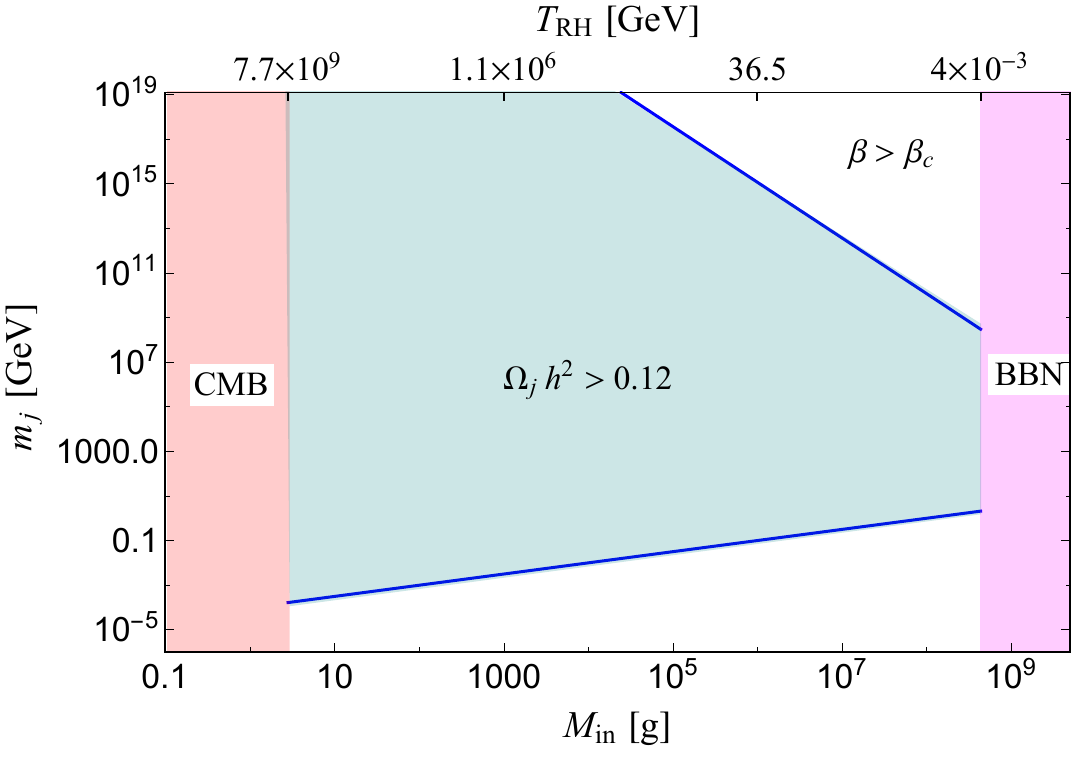}
     \caption{\it Allowed region in the ($M_{\rm in}\,,m_j$) and ($T_{\rm RH}\,,m_j$) plane  for $\beta >\beta_c$. 
    %This parameter space is independent of $(\beta,\,w_\phi)$. 
The blue lines indicate the correct relic of DM and the green-shaded region is excluded due to overproduction. Whereas the white regions are allowed but need any second mechanism to satisfy the relic. BBN and CMB bounds exclude the magenta and red-shaded regions, respectively; see text for details.
}
    \label{Fig:mdmvsmbh}
\end{figure}

Combining Eq.(\ref{Eq:omegah2}) with
Eqs.(\ref{Eq:ntotmjless}) and 
(\ref{Eq:nbhev}),
we obtained for $m_j <T_{\rm BH}^{\rm in}$

\bea
&&
\frac{\Omega_jh^2}{0.12}= 
4.2\times 10^7~\frac{g_0\,g_j}{\sqrt{g_*(T_{\rm BH})}g_{RH}^\frac{1}{4}}\sqrt{\frac{M_P}{\Min}}
\frac{m_j}{{\rm GeV}}
\nonumber
\\
&\simeq& 4.9\times10^6  \sqrt{\frac{M_P}{\Min}}
\frac{m_j}{\rm GeV}
\simeq
\sqrt{\frac{10^8\rm g}{\Min}}\frac{m_j}{1~\rm GeV}\,,
\label{Eq:omegah21}
\eea
where we took $g_{\rm RH}=g_*(T_{\rm BH})=106.75$ and DM is scalar.
We recognize from Eq.(\ref{Eq:omegah21}) the main feature we guessed in the introduction: for a given DM
mass $m_j$, the relic abundance {\it decreases} for 
increasing values of $\Min$ due to the lack of efficiency in the decay rate $\Gamma_{\rm BH}$, Eq.(\ref{Eq:gammabh}). 
We also obtained the interesting result that for a 1 GeV dark matter, the right relic abundance is obtained for reasonable PBH masses of 
$\Min \sim 10^8$ g.
 We illustrate our results in Fig.(\ref{Fig:mdmvsmbh}), 
where we recognize the slope $m_j \propto \sqrt{\Min}$ in the bottom blue solid line in the plot,
 corresponding to the dependence obtained in Eq.(\ref{Eq:omegah21})
 to achieve the right relic abundance. On the left of this line, too efficient PBH decay excludes a large region of the parameter space due to the overclosure of the Universe shown in greed-shaded 
 region. The pink-shaded regions 
 indicate the forbidden window of PBH mass, which 
 can disturb the BBN phase by introducing extra 
 relativistic degrees of freedom. The minimum PBH mass obtained from the brown-shaded region is set 
 by the maximum energy scale of inflation, which is 
 constrained by the CMB observation.

For $m_j > \tbh^{\rm in}$, the same exercise, using
Eq.(\ref{Eq:ntotmjmore}) instead of (\ref{Eq:ntotmjless}) gives
\bea
&&
\frac{\Omega_j h^2}{0.12}=
4.2\times 10^7~\frac{g_0\,g_j}{\sqrt{g_*(T_{\rm BH})}g_{RH}^\frac{1}{4}}
\frac{M_P^{\frac{9}{2}}}{\Min^{\frac{5}{2}}m_j^2}
\frac{m_j}{{\rm GeV}}\,,
\nonumber
\\
&&\simeq
4.9\times 10^6 \frac{M_P^{\frac{9}{2}}}{\Min^{\frac{5}{2}}m_j^2}
\frac{m_j}{\rm GeV}\,,
\label{Eq:omegah22}
\\
&&
\simeq \left(\frac{10^8 \rm g}{\Min}\right)^\frac{5}{2}\left(\frac{1.1\times10^{10} \rm GeV}{m_j}\right)\,.
\eea
We also recover the features we guessed in the introduction. For a given PBH population of mass $\Min$,
the relic abundance {\it decreases} with $m_j$
as the decay efficiency is largely reduced for heavy dark matter candidates, as it is clear from Eq.(\ref{Eq:ntotmjmore}). We also observe this behavior on
the top right of Fig.(\ref{Fig:mdmvsmbh}), where a viable
region arises for large dark matter masses, following $m_j\propto \Min^{-{5}/{2}}$ as expected for a fixed relic abundance.
One of the main results of our work is then that, in the case of PBH domination, the constraints on relic abundance allow only two very distinct regions: $10^{-5} ~\mbox{GeV} \lesssim m_j \lesssim 1$ GeV (corresponding to
$ m_j \ll T_{\rm BH}$), for which the upper and lower limits are set by BBN and CMB respectively. On the other hand, for $m_j \gg T_{\rm BH}$, we obtained $M_P\gtrsim m_j \gtrsim 10^8$ GeV, for which the lower limit is set by the BBN, and the upper limit is essentially the maximum possible mass that a fundamental particle can possibly possess namely the Planck mass.

\subsubsection{$\beta<\beta_c$}

\begin{figure*}[t!]
    \centering
    \includegraphics[scale=0.44]{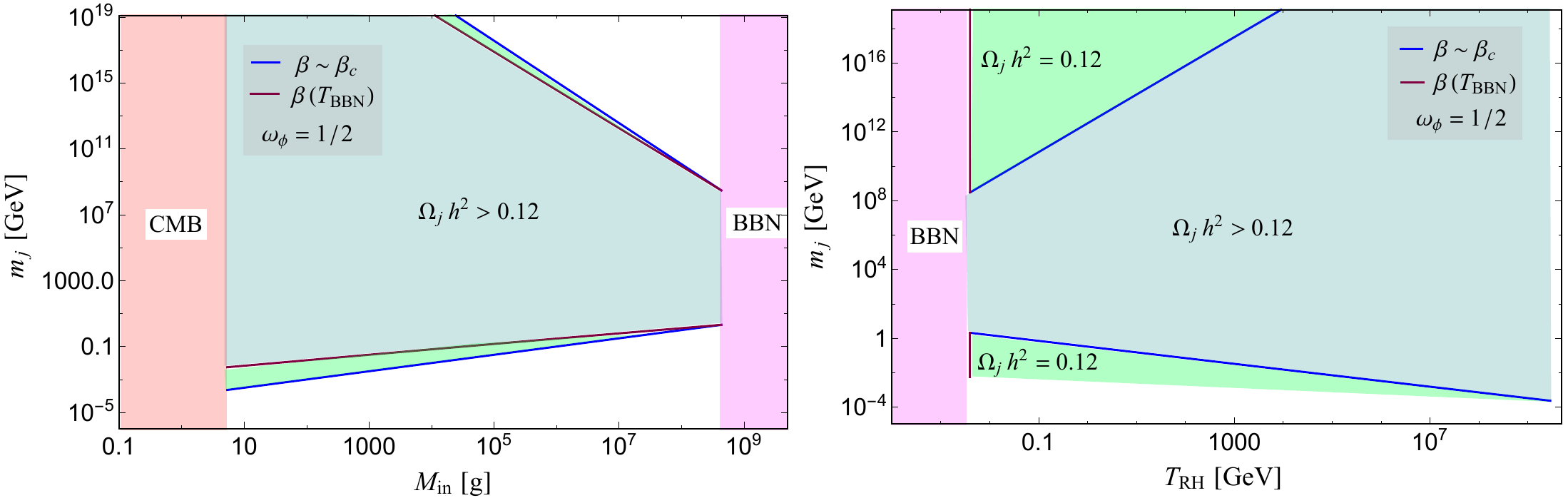}~~ \caption{\it Same as Fig.(\ref{Fig:mdmvsmbh}) 
    for $\beta \leq\beta_c$.  Here, the light green region 
    indicates the allowed parameter space in the context of PBH reheating, with an inflaton coupling below the critical value $y_\phi \ll y_\phi^c$. We show two limiting values of $\beta$: the critical value $\beta_{\rm c}$ in blue below which the inflaton dominate the energy budget {\it before} the PBH decay and $\beta(T_{\rm BBN}) < \beta_c$ in red which is the value of $\beta$ for which the reheating temperature due to PBHs decay is the BBN bound $T_{\rm BBN}=4 \,\rm MeV$. The deep green region indicates DM overproduction, whereas the white region indicates underproduction, where another mechanism to satisfy the correct DM abundance is required.}
    \label{fig:dmbetalbetac}
\end{figure*}

If $\beta < \beta_c$, the PBHs decay {\it during} inflaton $\phi$ domination, and PBHs never dominate the energy budget of the Universe. 
However, for $w_\phi> 1/3$ and\footnote{The condition $w_\phi>1/3$ ensures 
that the inflaton redshifts faster than the radiation.} inflaton coupling below some critical value $y_\phi^c$, PBH evaporation still determines the reheating temperature $\trh$ \cite{RiajulHaque:2023cqe}. 
Writing the coupling between the inflaton and the radiation under the form $y_\phi\phi\bar{f}f$, one can estimate the expression for the critical coupling $y_\phi^c$ (given in the appendix-\ref{appendixA}).
\iffalse
for $n<7$ ($w_\phi<\frac{5}{9}$) as\footnote{For a detailed derivation, see Appendix-\ref{appendixA}.}
\bea \label{Eq:criticalcoup}
&&
y_\phi^c=\sqrt{\frac{8\pi}{\alpha_n}}\,\beta^{\frac{3\,(1-w_\phi)}{4\,(3w_\phi-1)}}\left(\frac{48\pi^2}{\lambda}\right)^{\frac{1-w_\phi}{4\,(1+w_\phi)}}
\\
&&
\left(\frac{\epsilon\,\gamma^{-3\,w_\phi}}{2\pi\,(1+w_\phi)}\right)^{\frac{1-w_\phi}{2\,(1-3\,w_\phi)\,(1+w_\phi)}}\left(\frac{M_P}{\Min}\right)^{\frac{3}{2}\frac{(1-w_\phi)^2}{(1-3w_\phi)\,(1+w_\phi)}}\,,
\nonumber
\eea
 where $\alpha_n=\frac{2\,(1+w_\phi)}{(5-9\,w_\phi)}\sqrt{\frac{6\,(1+w_\phi)\,(1+3w_\phi)}{(1-w_\phi)^2}}$  and $\lambda$ is the potential parameter defined in the Appendix-\ref{appendixA} {\blue \bf [not yet defined in appendix A]}. However for $n>7$ ($w_\phi>\frac{5}{9}$) the above critical coupling takes the following form 
\bea \label{Eq:criticalcoup1}
&&
y_\phi^c=\sqrt{-\frac{8\pi\,\beta}{\alpha_n}}\,(48\pi^2)^{\frac{3\,w_\phi-1}{6\,(1+w_\phi)}}\,\lambda^{\frac{w_\phi-1}{4\,(1+\,w_\phi)}}
\\
&&
\left(\frac{\epsilon\,\gamma^{-3\,w_\phi}}{2\pi\,(1+w_\phi)}\right)^{-\frac{1}{3\,(1+w_\phi)}}\left(\frac{M_P}{\Min}\right)^{\frac{1-w_\phi}{1+w_\phi}}\left(\frac{\rho_{\rm end}}{M_P^4}\right)^{\frac{5-9\,w_\phi}{12\,(1+\,w_\phi)}}\,,
\nonumber
\eea
where $\rho_{\rm end}$ is the inflaton energy density at the end of the inflation{\red \bf [Donald, can you check that too?]}. 
\fi
Similarly, once we fixed a particular coupling $y_\phi$,
%using the same expression Eq.(\ref{Eq:criticalcoup}) and (\ref{Eq:criticalcoup1}), 
there always exists a $\beta$ value $\beta_{\rm BH}$ above which the PBHs determine 
the reheating temperature\footnote{For very strong coupling $y_\phi$ such that inflaton completely decays before the completion of the evaporation process, we always required PBH domination ($\beta_{\rm BH}\sim \beta _c$) to ensure PBH reheating.}.

However, there is a little subtlety here. The 
temperature generated by the PBH evaporation, $T_{\rm ev}$, at
$\aev$ is not strictly speaking the reheating temperature
$\trh$. \iffalse
Despite its faster dilution, inflaton still dominates the Universe either until $t_\phi=\Gamma_\phi^{-1}$,
or when $\rho_\phi<\rho_R$ due to the lesser rate of radiation production both from PBHs and inflation.
\fi However, as stated earlier, the Yukawa is too small ($y \ll y_\phi^c$) to affect the temperature of the thermal bath and $T$ being reshifted from $t_{\rm ev}$ to
$t_{\rm rh}$ (time scale when $\rho_\phi=\rho_R$). Since dark matter and radiation production are concluded at the time of PBH evaporation,  instead of  Eq.(\ref{Eq:omegah2}), it is useful to use the following expression for the abundance, 
\beq
\Omega_jh^2= 1.6\times 10^8\,\frac{g_0}{g_{\rm RH}}\frac{N_j\times n_{\rm BH}(\aev)}{T_{\rm ev}^3}\,\frac{m_j}{\text{GeV}}\,,
\label{Eq:omegah2m}
\eeq
because the dominant temperature provided by the PBH decay 
follow a (quasi) iso-entropic law $T \propto a^{-1}$  between $\aev$ and $\arh$.

The energy density of PBHs being transferred to the radiation at the evaporation time, one has $\rho_{\rm R}(\aev)=\rho_{\rm BH}(\aev)$,
or
\beq
\frac{n_{\rm BH}(\aev)}{T_{\rm ev}^3}=\left(\frac{\alpha_T^3\rho_{\rm BH}(\aev)}{\Min^4}\right)^\frac{1}{4}\,.
\label{Eq:nt3}
\eeq
 Implementing Eq.(\ref{Eq:nt3})
in Eq.(\ref{Eq:omegah2m}) using (\ref{Eq:bhenergyden}) and
(\ref{Eq:ainaev}) one obtains for $m_j < T_{\rm BH}^{\rm in}$

\begin{figure*}[t!]
    \centering
    \includegraphics[scale=0.45]{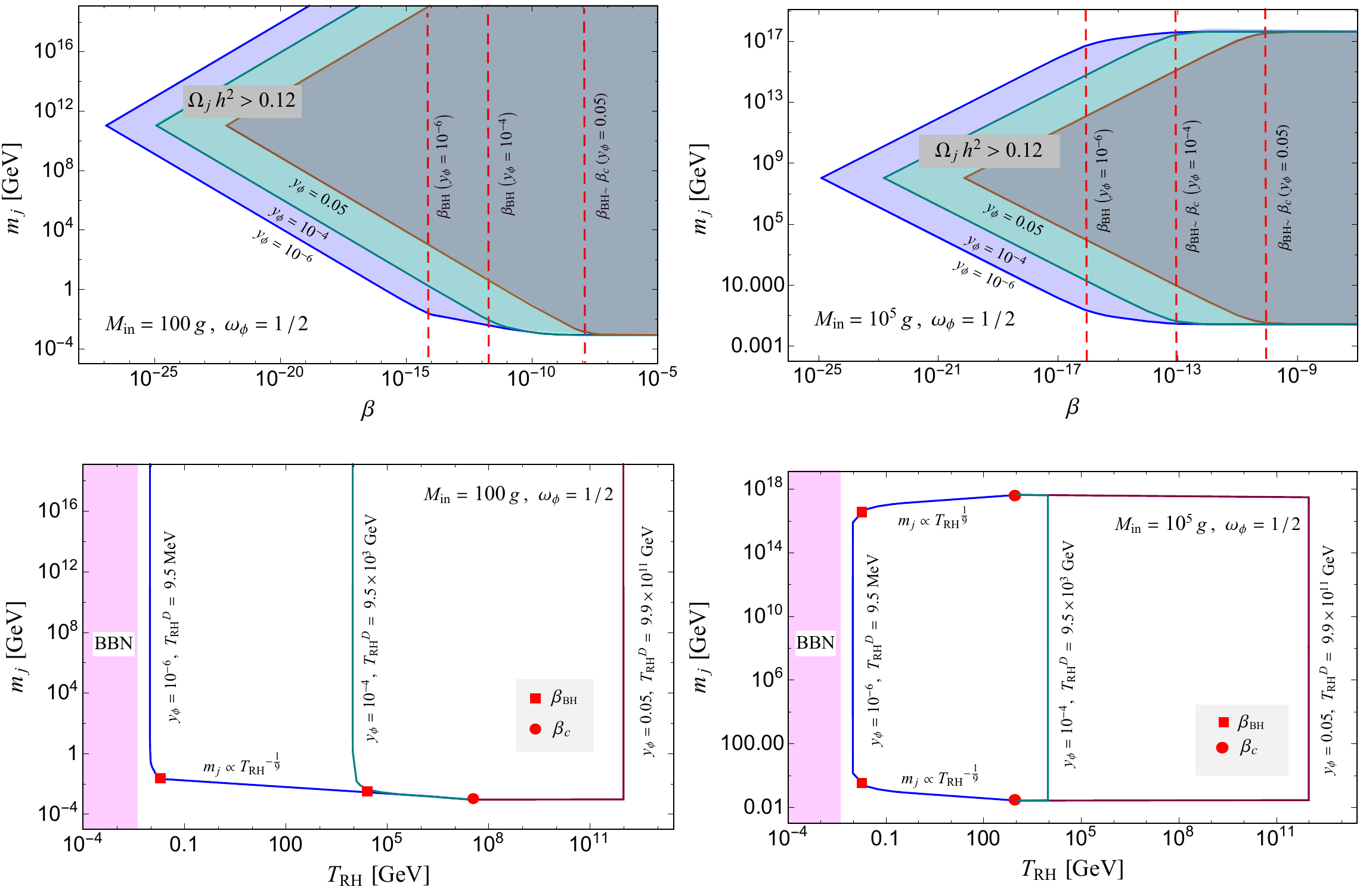}~~ \caption{\it Allowed parameter space for $w_\phi=1/2$ in the ($m_j$,\,$\beta$) and ($m_j$,\,$T_{\rm RH}$) plane  for two different values of $\Min=(10^2,\,10^5)$ g and three distinct values of $y_\phi=(0.05,\,10^{-4},10^{-6})$. The shaded region in the upper panel plots indicates the forbidden regime due to overproduction. The vertical red dashed line indicates the divider line between the inflaton and PBH reheating. The regime on the right-hand side of the red dashed line is dominated by PBH reheating, whereas on the left-hand side, the Yukawa coupling $y_\phi$ defines the reheating temperature.}
    \label{fig:inflatonreheatingn6}
\end{figure*}

\beq
\frac{\Omega_jh^2}{0.12}=
2.8\times 10^{8}~ \mu_j\beta^{\frac{1}{4}}\frac{g_0g_j}{g_{\rm RH}^{\frac{1}{4}}\,g_*(T_{\rm BH})}
\left(
\frac{M_P}{\Min}
\right)^\frac{1-w_\phi}{2+2w_\phi}
\frac{m_j}{\rm{GeV}},
\label{Eq:omegah2inflatonsmallmj}
\eeq

with

\beq
\mu_j=\left(\frac{\epsilon \pi^{w_\phi}\gamma^{w_\phi}}{2 + 2 w_\phi}\right)^{\frac{1}{2+2w_\phi}}
\eeq

\noindent
whereas
\beq
\frac{\Omega_jh^2}{0.12}=
1.7\times 10^{45} \mu_j\beta^{\frac{1}{4}}\frac{g_0g_j}{g_{\rm RH}^{\frac{1}{4}}g_*(T_{\rm BH})}
\left(
\frac{M_P}{\Min}
\right)^\frac{5+3w_\phi}{2+2w_\phi}\frac{\rm GeV}{m_j}\,.
\label{Eq:omegah2inflatonlargemj}
\eeq
for $m_j >T_{\rm BH}^{\rm in}$. Note that we find similar expressions as Eqs.(\ref{Eq:omegah21}) and (\ref{Eq:omegah22}) if one 
sets $w_\phi=0$. This is expected as for $w_\phi=0$, the inflaton field behaves like dust, as PBH does.

We illustrate our result in Figs.(\ref{fig:dmbetalbetac}), where we
show the same parameter space as in the case of PBHs domination Fig.(\ref{Fig:mdmvsmbh}). 
To make the comparison easier,
we have chosen two extreme parameters, 
$\beta=\beta_c$, corresponding to the preceding case, and $\beta=\beta(T_{\rm BBN})$ corresponding to the case where the reheating temperature generated by
the PBHs decay corresponds to the BBN temperature, $T_{\rm BBN}\sim 4$ MeV.

It is clear from Eqs.(\ref{Eq:omegah2inflatonsmallmj})
and (\ref{Eq:omegah2inflatonlargemj})
that in this case, a new region of allowed parameter space opens up.
Indeed, contrary to PBH 
domination, where only the 
lifetime of PBH (and not $\beta$) determined the relic abundance; here, along with the lifetime, the initial fraction also controls the abundance $\Omega_{j}h^2 \propto \beta^\frac{1}{4}$. In other
words, for a given $m_j$, lowering $\beta$ lowers the relic abundance, opening new allowed regions compared to the case $\beta\gtrsim \beta_c$. 
We represent these new regions in light green, for $w_\phi=\frac{1}{2}$
on the left panel of Fig.(\ref{fig:dmbetalbetac}). Whereas the deep green region still overclose the Universe, the regions
between the lines $\beta=\beta(T_{\rm BBN})$ and $\beta=\beta_c$ possess the 
right relic abundance. 

The slopes are given by Eq.(\ref{Eq:omegah2inflatonsmallmj})
and (\ref{Eq:omegah2inflatonlargemj}), 
i.e. $m_j \propto \beta^{-\frac{1}{4}}\,\Min^{\frac{1-w_\phi}{2\,(1+w_\phi)}}$ for $m_j\ll T_{\rm BH}^{\rm in}$, and
$m_j \propto \beta^{\frac{1}{4}}\Min^{-\frac{5+3w_\phi}{2\,(1+w_\phi)}}$ for $m_j \gg T_{\rm BH}^{\rm in}$. However, the reheating being
completed by the PBHs population, the reheating temperature can also be expressed in terms of the fraction $\beta$ as,
\beq
\trh\sim M_P\beta^{\frac{3}{4}\frac{1+w_\phi}{3w_\phi-1}}
\left(\frac{\Min}{M_P}\right)^{\frac{3}{2}\frac{1-w_\phi}{3w_\phi-1}}
\label{Eq:trhpbh}\,,
\eeq
as shown in \cite{RiajulHaque:2023cqe}, also detailed in the Eq.(\ref{Eq:pbhreheattemp}) of our dedicated appendix. Therefore, for a fixed value of reheating temperature, $\beta$ behaves as proportional to $\Min^{-\frac{2\,(1-w_\phi)}{1+w_\phi}}$. Thus, $m_j\propto \Min^\frac{1-w_\phi}{1+w_\phi}$
for $m_j \ll T_{\rm BH}^{\rm in}$, and
$m_j\propto\Min^{-\frac{3+w_\phi}{1+w_\phi}}$ for $m_j \gg T_{\rm BH}^{\rm in}$.
For $w_\phi=\frac{1}{2}$, this corresponds to slopes 
$m_j \propto \Min^{\frac{1}{3}}$ for $m_j\ll T_{\rm BH}^{\rm in}$, and
$m_j \propto \Min^{-\frac{7}{3}}$ for $m_j \gg T_{\rm BH}^{\rm in}$, which is effectively what is observed on the red lines in the left panel of Fig.(\ref{fig:dmbetalbetac}) where we fixed
the reheating temperature to be $\trh=T_{\rm BBN}$.

We show in the right panel of Fig.(\ref{fig:dmbetalbetac}) the same analysis but in the ($\trh\,,m_j$)-plane. 
This is just another
representation of our results.
Indeed, 
combining Eqs.(\ref{Eq:trhpbh}) and (\ref{Eq:omegah2inflatonsmallmj})
or (\ref{Eq:omegah2inflatonlargemj}),
for each couple ($\trh\,,m_j$), there exists a unique couple ($\beta\,,\Min$) which fixes
$\trh$ and $\Omega_jh^2$. 
We obtain $m_j \propto \trh^{\frac{3w_\phi-1}{3\,(1+w_\phi)}}$ for a fixed value of $\beta<\beta_c$ when $m_j\ll T_{\rm BH}^{\rm in}$ and $m_j \propto \trh^{-\frac{(3w_\phi-1)\,(5+3w_\phi)}{(1+w_\phi)\,(1-w_\phi)}}$ for  $m_j\gg T_{\rm BH}^{\rm in}$. For example, for $w_\phi=\frac{1}{2}$ and any fixed value of $\beta$, $m_j\propto \trh^{\frac{1}{9}}$ for $m_j\ll T_{\rm BH}^{\rm in}$ and $m_j\propto \trh^{-\frac{13}{3}}$ for $m_j\gg T_{\rm BH}^{\rm in}$. 
Similarly, for a fixed $\Min$, we have $m_j \propto \trh^{\frac{1-3w_\phi}{3\,(1+w_\phi)}}$ for $m_j\ll T_{\rm BH}^{\rm in}$ and $m_j \propto \trh^{\frac{3w_\phi-1}{3\,(1+w_\phi)}}$ for  $m_j\gg T_{\rm BH}^{\rm in}$, which gives for $w_\phi=\frac{1}{2}$, $m_j \propto \trh^{-\frac{1}{9}}$ for $m_j\ll T_{\rm BH}^{\rm in}$ and $m_j \propto \trh^{\frac{1}{9}}$ for  $m_j\gg T_{\rm BH}^{\rm in}$ (see, for instance, lower panel of Fig.(\ref{fig:inflatonreheatingn6})). \\

Note that PBHs can emit inflaton particles. Accounting for the production of the inflaton particle from evaporation modifies the total number of degrees of freedom $g_*(T_{\rm BH})$ at $T_{\rm BH}$ associated with the evaporation function, i.e,   $\epsilon = \frac{27}{4} \frac{g_*(T_{\rm BH})\, \pi}{480}$. Now the total degrees of freedom account for the standard model particles, dark matter, and inflaton. Taking $g_{\rm SM} = 106.75$ and $g_{\phi} = 1$, it can be inferred that the production of inflaton particles from evaporation would amount to approximately 1\% of that of radiation, which we can safely ignore compare to the thermal bath. Moreover, we are mainly interested in the scenario where the inflaton equation of state $w_\phi\geq1/3$, so it redshifts equal to or faster than radiation and faster than dark matter. Therefore, even at the evaporation end, the inflaton number density is comparable to the dark matter but with its subsequent evolution becomes negligible compared to the dark matter abundance.

\

\subsection{Inflaton reheating}

If $\beta< \beta_c$ and the inflaton coupling 
is strong enough, $y_\phi \gtrsim y_\phi^c$, the 
inflaton dominates the reheating process {\it and
also} determines 
$\trh$. 
However, depending on the coupling strength, the reheating, which is governed by the inflaton decay width 
$\Gamma_\phi$ may happen before or after the evaporation point $\aev$. 
We give in the appendix-\ref{appendixb} the threshold
value $y_\phi^{\rm th}$ 
above which the inflaton decays before the PBH evaporates as a function of the parameters of the inflationary potential $V(\phi)$.

Note that in our analysis, we always supposed 
that the PBHs are formed {\it during} the reheating, which means $\ain<\arh$.
This also means that there exists, for each 
PBH mass range, an {\it upper} size of the horizon to ensure their formation during the reheating period. This upper size of the horizon can be converted into a maximum reheating temperature.  
This temperature can be estimated from $H_{\rm in}$,
 $T_{\rm RH}^{\rm max}=\sqrt{\frac{\sqrt{3}\,H_{\rm in}\,M_P}{\alpha_{\rm T}^{1/2}}}=\left(\frac{\gamma}{0.2}\right)^{1/2}\,\left(\frac{1.9\times 10^3\,\rm g}{M_{\rm in}}\right)^{1/2}10^{14}\,\rm GeV$ (see, for instance Eq.(\ref{Eq:min})). As an example for PBH mass of $10^{5}$ g,  $T_{\rm RH}^{\rm max}\sim 1.4\times 10^{13}$ GeV, and $T_{\rm RH}^{\rm max}\sim 1.4\times 10^{11}$ GeV for $M_{\rm in}=10^9$ g.\footnote{We thank the referee for having pointed out this remark.}

Following the same procedure as in the previous section, we will discuss in detail the two possibilities.
\subsubsection{$\aev<\arh$}
 If $y_\phi^{\rm c}<y_\phi<y_\phi^{\rm th}$, the PBHs evaporate before 
the end of reheating thus, $\aev<\arh$. The dilution of the dark component
between $\aev$ and $\arh$ is then modified compared
to our previous analysis due the injection 
of a considerable amount of entropy during the decay
of the inflaton, which was negligible in the previous section.
This injection can considerably dilute the relic abundance, and we expect an increase in the allowed mass range, allowing heavier dark matter. 
Whereas $N_j$ is not modified by the presence of the inflaton, $n_{\rm BH} (\aev) \left({\aev}/{\arh}\right)^3$ appearing in Eq.(\ref{Eq:omegah2})
is affected. 
Connecting the evolution of the scale factor from the evaporation point to the reheating time in the inflaton-dominated era, we obtain 
\bea
&&
\frac{\aev}{\arh}=
\left(\frac{t_{\rm ev}}{t_{\rm RH}}\right)^\frac{2}{3(1+w_\phi)}
=\left(\frac{3(1+w_\phi)}{2}\frac{H_{\rm RH}}{\Gamma_{\rm BH}}\right)^\frac{2}{3(1+w_\phi)}
\nonumber
\\
&&
%=\left(\frac{\sqrt{\rho_{\rm RH}}\Min^3}{\delta M_P^4}\,,
=\left(\frac{(1+w_\phi)}{2\sqrt{3}}\frac{\sqrt{\alpha_T}\trh^2}{M_P}\frac{\Min^3}{\epsilon M_P^4}\right)^\frac{2}{3(1+w_\phi)}\,.
\label{Eq:aevarh1}
\eea
The number density of the species $j$ at $\arh$ is then given by Eq.(\ref{Eq:nbhaev}) modulo 
the dilution factor between $\aev$ and $\arh$
\beq
n_j(\arh)=N_j\times n_{\rm BH}(\aev)\times 
\left(\frac{\sqrt{\alpha_T}\,(1+w_\phi)}{2\sqrt{3} \,\epsilon}\frac{\trh^2\Min^3}{M_P^5}\right)^{\frac{2}{1+w_\phi}}.
\label{Eq:ndmarh}
\eeq
Utilizing Eq.(\ref{Eq:ndmarh}) one obtains the ratio
\beq \label{Eq:arhgaev}
\frac{n_j(\arh)}{\trh^3}=\tilde \mu\left( \frac{M_P^{4w_\phi-2}\Min^{1-w_\phi}}{\trh^{3w_\phi-1}}\right)^\frac{1}{1+w_\phi}
\,,
\eeq
for $m_j<\tbh^{\rm in}$
with

\beq \label{Eq:arhggaev}
\tilde \mu = \frac{720\, g_j\,\zeta(3)\,\beta}{g_*(T_{\rm BH})\,\pi^2}
\left(\frac{\sqrt{\alpha_T}\,\gamma^{w_\phi}}{4\sqrt{3}\,\pi}\right)^\frac{2}{1+w_\phi}\,.
\eeq

This gives for $g_{\rm RH}=g_*(T_{\rm BH})=106.75$,

\bea
&&
\frac{\Omega_j h^2}{0.12} =4.9 \times 10^7\tilde \mu
\left( \frac{M_P^{4w_\phi-2}\Min^{1-w_\phi}}{\trh^{3w_\phi-1}}\right)^\frac{1}{1+w_\phi}\frac{m_j}{1~\rm GeV}
\nonumber
\\
&&
\simeq3.4\times 10^6 \beta \left( \frac{M_P^{4w_\phi-2}\Min^{1-w_\phi}}{\trh^{3w_\phi-1}}\right)^\frac{1}{1+w_\phi}\frac{m_j}{1~\rm GeV}
\label{Eq:omegah2inflatonapprox1}
\\
&&
\simeq3.4\times 10^6 \beta \left(\frac{\Min}{M_P}\right)^{\frac{1-w_\phi}{1+w_\phi}}\left(\frac{M_P}{T_{\rm RH}}\right)^{\frac{3\,w_\phi-1}{1+w_\phi}}\frac{m_j}{1~\rm GeV}
\nonumber
\,.
\eea

For $m_j>\tbh^{\rm in}$, we obtain

\beq \label{Eq:arhgaev2}
\frac{n_j(\arh)}{\trh^3}=\frac{\tilde \mu}{m_j^2}\left( \frac{M_P^{2+8w_\phi}}{\Min^{1+3w_\phi}\trh^{3w_\phi-1}}\right)^\frac{1}{1+w_\phi}
\,,
\eeq
and

\bea
&&
\frac{\Omega_j h^2}{0.12} =4.9 \times 10^7
\frac{\tilde \mu}{m_j^2}\left( \frac{M_P^{2+8w_\phi}}{\Min^{1+3w_\phi}\trh^{3w_\phi-1}}\right)^\frac{1}{1+w_\phi}\frac{m_j}{1~\rm GeV}
\nonumber
\\
&&
\simeq3.4\times 10^6 \frac{\beta}{m_j^2}\left( \frac{M_P^{2+8w_\phi}}{\Min^{1+3w_\phi}\trh^{3w_\phi-1}}\right)^\frac{1}{1+w_\phi}\frac{m_j}{1~\rm GeV}
\label{Eq:omegah2inflatonapprox2}
\\
&&
\simeq2\times 10^{30} \beta \left(\frac{M_p}{\Min}\right)^{\frac{1+3\,w_\phi}{1+w_\phi}}\left(\frac{M_P}{T_{\rm RH}}\right)^{\frac{3\,w_\phi-1}{1+w_\phi}}\,\frac{10^{13}\,\rm GeV}{m_j}\,.
\nonumber
\eea

\begin{figure}
    \centering
    \includegraphics[scale=0.42]{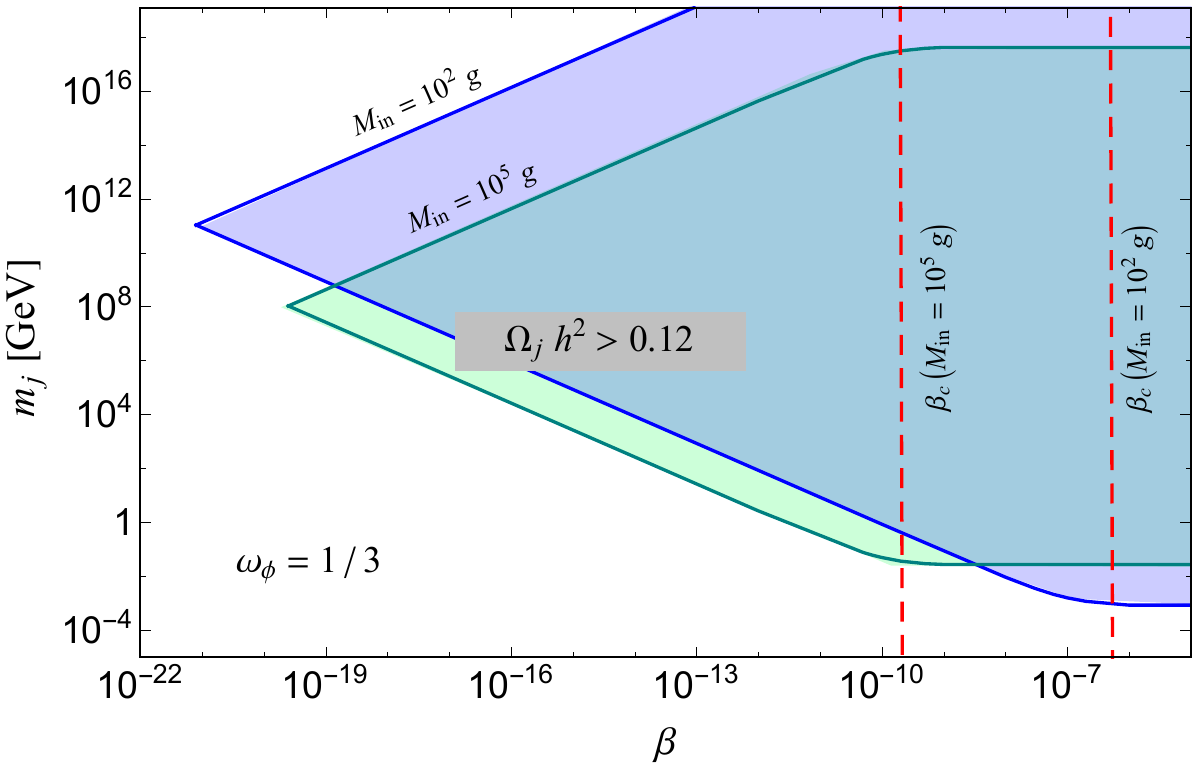}~~ \caption{\it \it Allowed parameter space for $w_\phi=1/3$ in the ($m_j$,\,$\beta$) plane  for two different values of $\Min=(10^2,\,10^5)$ g. Interestingly,  once we fixed $M_{\rm in}$, $(m_j,\,\beta)$ parameter space turns out to be independent of coupling value $y_\phi$.}
    \label{fig:inflatonreheatingn4}
\end{figure}

\begin{figure}
    \centering
    \includegraphics[scale=0.42]{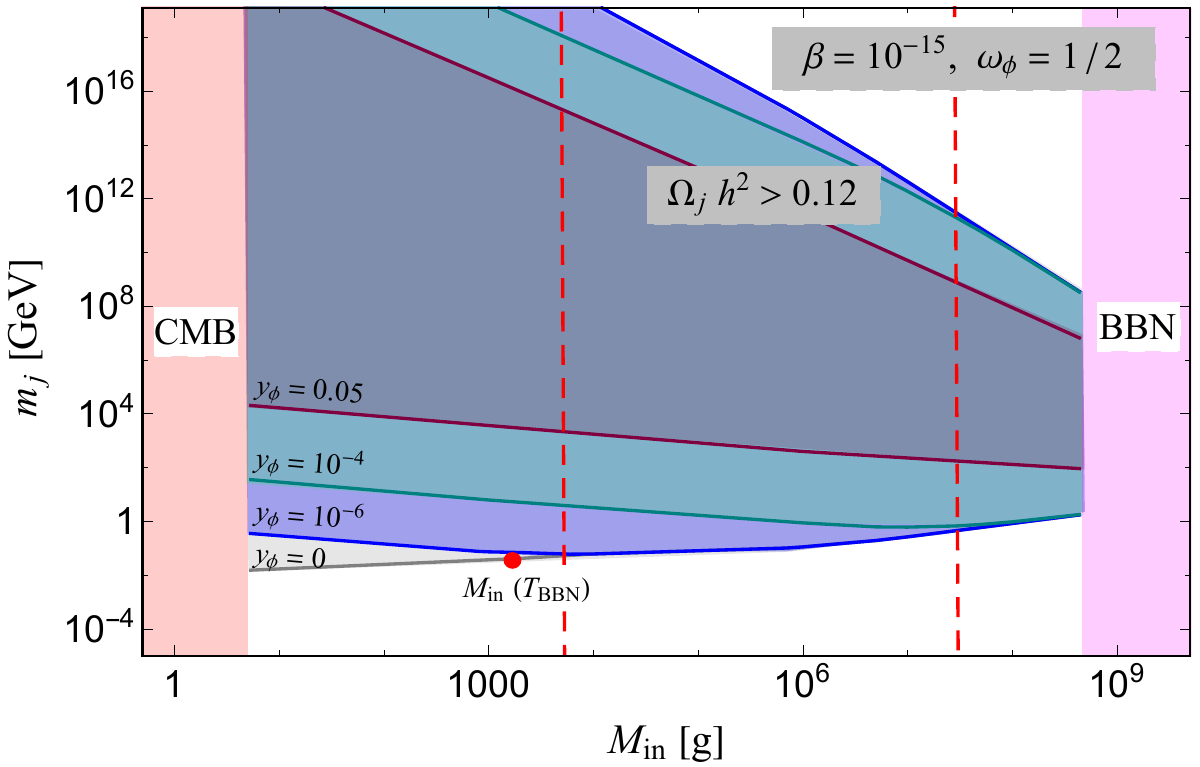}
     \caption{\it Allowed region in the ($M_{\rm in}\,,m_j$) plane for four different values of the coupling $y_\phi=(0.05,\,10^{-4},\,10^{-6},\,0)$ with $w_\phi=\frac{1}{2}$ and $\beta=10^{-15}$. 
    %This parameter space is independent of $(\beta,\,w_\phi)$. 
    BBN and CMB bounds exclude the magenta and red-shaded 
    regions, respectively. 
    For each value of 
    $y_\phi$, the 
    vertical red dashed lines (left one is for $y_\phi=10^{-6}$ and the right one is for $y_\phi=10^{-4}$) separate the regions 
    between the inflaton and PBH 
    reheating. The regions on the right side of the red 
    dashed lines are dominated by 
    PBH reheating, whereas on the 
    left side, the Yukawa  
    reheating temperature is
    determined by $y_\phi$. The red circle represents the PBH formation mass associated with the case where reheating temperature from the PBH decay corresponds to the BBN temperature and here we set coupling $y_\phi=0$. The shaded region indicates DM overproduction. Whereas the white region indicates under production, here we need any second mechanism to satisfy the correct DM relic.}
    \label{Fig:mdmvsmbh2}
\end{figure}
\subsubsection{$\aev>\arh$}
For even stronger inflaton coupling $y_\phi> y_\phi^{\rm th} > y_\phi^{\rm c}$, 
there exists the possibility that the inflaton decays even
{\it before} the completion of the PBH evaporation process.
In other words, $\aev>\arh$. 
In this case, we have
\bea\label{aevlessarh}
&&
\frac{n_j(\aev)}{T^3(\aev)}
=\frac{N_j\times n_{\rm BH}(\aev)}{T^3(\aev)}=N_j\frac{n_{\rm BH}(\ain)\left(\frac{\ain}{\aev}\right)^3}{T_{\rm RH}^3\left(\frac{\arh}{\aev}\right)^3}
\nonumber
\\
&&
=N_j\times \frac{48\pi^2\gamma^2\beta\frac{M_P^6}{\Min^3}\left(\frac{\ain}{\arh}\right)^3}{T_{\rm RH}^3}\,,
\eea
where we supposed that no particles had decoupled from the
thermal plasma between $\arh$ and $\aev$, and used
Eq.(\ref{Eq:rhophiin}). The scale factor between the formation and reheating point can be connected through the evolution of the Hubble parameter as,

\beq\label{Eq:ainarh}
\frac{\ain}{\arh}=\left(\frac{H_{\rm RH}}{H_{\rm in}}\right)^{\frac{2}{3\,(1+w_\phi)}}=\left(\sqrt{\frac{\alpha_T}{3}}\frac{\Min\,T_{\rm RH}^2}{4\pi\,\gamma\,M_P^3}\right)^{\frac{2}{3\,(1+w_\phi)}}\,,
\eeq
where $H_{\rm RH}$ is the Hubble parameter at the end of reheating.

Implementing Eq.(\ref{Eq:ainarh}) into Eq.(\ref{aevlessarh}), one can  find the ratio
\beq \label{Eq:arhgaev4}
\frac{n_j(\aev)}{T^3(\aev)}=\tilde \mu\left( \frac{M_P^{4w_\phi-2}\Min^{1-w_\phi}}{\trh^{3w_\phi-1}}\right)^\frac{1}{1+w_\phi}\,
\,,
\eeq 
for $m_j<T_{\rm BH}^{\rm in}$ and for $m_j>T_{\rm BH}^{\rm in}$
\beq \label{Eq:arhgaev3}
\frac{n_j(\aev)}{T^3(\aev)}=\frac{\tilde \mu}{m_j^2}\left( \frac{M_P^{2+8w_\phi}}{\Min^{1+3w_\phi}\trh^{3w_\phi-1}}\right)^\frac{1}{1+w_\phi}\,,
\eeq
Interestingly, the above equations (\ref{Eq:arhgaev4}) and (\ref{Eq:arhgaev3}) turn out to be exactly the same with the previous case $\arh>\aev$ (see, for instance, Eq.({\ref{Eq:arhgaev}) and (\ref{Eq:arhgaev2}) ). Therefore, the DM abundance naturally follows Eqs.($\ref{Eq:omegah2inflatonapprox1}$) and (\ref{Eq:omegah2inflatonapprox2}).}
An easier way to understand this is to notice that the present relic abundance is given by 
\beq
n_j(a_0)=n_j(\aev)
\left(\frac{\aev}{a_0}\right)^3=
n_{\rm BH}(\aev)N_j\left(\frac{\aev}{a_0}\right)^3\,,
\nonumber
\eeq
which gives
\bea
&&
n_j(a_0)=n_{\rm BH}(\ain)N_j\left(\frac{\ain}{a_0}\right)^3
\nonumber
\\
&&
=n_{\rm BH}(\ain)N_j\left(\frac{\ain}{\arh}\right)^3\left(\frac{\arh}{a_0}\right)^3
\,.
\nonumber
\eea
If the dilution is dominated by the same field (in this case the inflaton) between $\ain$ and $\arh$, the relic abundance does not depend on the evaporation time.

\iffalse

{\red Implementing Eq.(\ref{Eq:ainarh}) into Eq.(\ref{aevlessarh}), one finds for the dark matter abundance, Eq.(\ref{Eq:omegah2m}) :

\beq
\Omega_jh^2=5.9\times10^6\frac{N_jm_j}{1\rm GeV}\frac{48\,\pi^2\gamma^2\beta}{T_{\rm RH}^3}\frac{M_P^6}{\Min^3}\left(\sqrt{\frac{\alpha_T}{3}}\frac{\Min\,T_{\rm RH}^2}{4\pi\,\gamma\,M_P^3}\right)^{\frac{2}{(1+w_\phi)}}\,,
\eeq
For $m_j<T_{\rm BH}^{\rm in}$, the abundance is
then
\bea
&&
\frac{\Omega_jh^2}{0.12}\simeq 2.5\times 10^{13}\beta\left(\frac{10^{12}\,\rm GeV}{T_{\rm RH}}\right)^3\left(\frac{10^8\rm g}{\Min}\right)
\nonumber
\\
&&
\times\left[\gamma^{w_\phi}\left(\frac{T_{\rm RH}}{10^{12}\,\rm GeV}\right)^2\,\frac{\Min}{10^8\,\rm g}\right]^{\frac{2}{1+w_\phi}}\,\frac{m_j}{1\,\rm GeV}\,,
\label{Eq:omegah2smallmjbis}
\eea
whereas for for $m_j>T_{\rm BH}^{\rm in}$,
\bea
&&
\frac{\Omega_jh^2}{0.12}\simeq 2.8\times 10^{13}\beta\left(\frac{10^{12}\,\rm GeV}{T_{\rm RH}}\right)^3\left(\frac{10^8\,\rm g}{\Min}\right)^3
\nonumber
\\
&&
\times\left[\gamma^{w_\phi}\left(\frac{T_{\rm RH}}{10^{12}\,\rm GeV}\right)^2\,\frac{\Min}{10^8\,\rm g}\right]^{\frac{2}{1+w_\phi}}\frac{10^{10}\,\rm GeV}{m_j}\,.
\label{Eq:omegah2largemjbis}
\eea
}
{\magenta \bf The above red portion is not needed.}

\fi

We show in the upper panel of Fig.(\ref{fig:inflatonreheatingn6}) the allowed regions in the ($\beta\,,m_j$)
parameter space for different values of
$y_\phi$ and $\Min=100$ g (left) 
or $10^5$ g (right) with $w_\phi=\frac{1}{2}$. For each $y_\phi$, 
we also plotted
in the same figure the lines for $\beta_{\rm BH}$, corresponding to the value of $\beta$
above which the PBHs dominate the reheating process over the inflaton.
On the left side of the $\beta_{\BH}$
line, the inflaton reheats the Universe, and we recover the behavior
we found in Eqns.(\ref{Eq:omegah2inflatonapprox1}) and (\ref{Eq:omegah2inflatonapprox2}) i.e.
points allowed by the relic density constraint respects 
$m_j \propto \frac{1}{\beta}$
for $m_j \ll T_{\rm BH}^{\rm in}$ and 
$m_j \propto \beta$ for $m_j \gg T_{\rm BH}^{\rm in}$. Once $\beta>\beta_{\rm BH}$, $m_j$
follows the law $m_j\propto \beta^{-\frac{1}{4}}$ for $m_j\ll T_{\rm BH}^{\rm in}$ ($\beta^{\frac{1}{4}}$ for $m_j \gg T_{\rm BH}^{\rm in}$ )
as expected from Eq.(\ref{Eq:omegah2inflatonsmallmj}) and (\ref{Eq:omegah2inflatonlargemj}).
Then, once $\beta\gtrsim \beta_c$, the 
relic density depends only on the PBH lifetime and is then independent on $\beta$,
as we also noticed on Eq.(\ref{Eq:omegah21}).
We also show our result in the plane ($\trh,m_j$) in the lower 
panel. To this end, it may indeed be worth pointing the special case 
at $w_{\phi} =1/3$, for which DM abundance turns independent of $T_{\rm RH}$ or inflaton coupling $y_{\phi}$. For such case, also we 
showed the behavior in $(\beta, m_j)$ plane in Fig.(\ref{fig:inflatonreheatingn4}). 
Due to the indistinguishable nature between the inflaton and 
radiation, the intermediate $m_j\propto \beta^{\pm 1/4}$ behavior corresponding to PBH 
reheating does not arise, and hence $\beta_{\rm BH} =\beta_c$ 
condition satisfies as expected. 

Finally, for clear comparison with the two preceding cases (PBH reheating and domination, PBH reheating and inflaton domination), we plotted
in Fig.(\ref{Fig:mdmvsmbh2}) the
allowed region in the ($M_{\rm in},m_j$) plane for $w_\phi=\frac{1}{2}$ and different values of $y_\phi$. We clearly see that increasing $y_\phi$ exclude region shrinks naturally.
Indeed, for inflaton domination, the reheating temperature increases with $y_\phi$. As a consequence,  the same amount of relic abundance is obtained for {\it higher} dark
matter mass in the case $m_j < T_{\rm BH}^{\rm in}$ as one can 
see from Eq.(\ref{Eq:omegah2inflatonapprox1}).
For $m_j> T_{\rm BH}^{\rm in}$,
it is the opposite, see Eq.(\ref{Eq:omegah2inflatonapprox2}),
and a {\it lower} DM mass is necessary to obtain the right amount of relic abundance.
This is easy to understand, as higher reheating temperatures there is a tendency
to dilute the dark matter abundance more.
What is also interesting in this plot is the different regimes that can be observed for $M_{\rm in}$. For very large values of $M_{\rm in}$, 
not far from the BBN limit,
the reheating is determined by the PBH while they dominate the Universe, and we recover the results showed in Fig.(\ref{Fig:mdmvsmbh}).
For intermediate values of $M_{\rm in}$, the PBHs do not dominate the Universe but dominate the reheating process, whereas for low
$M_{\rm in}$, 
on the left side of the dashed lines, the PBHs dominate neither 
the Universe's energy budget nor the reheating process.
In this case, the dependence
between $M_{\rm in}$ and $m_j$
changes slope between these two regimes, as we can see 
comparing Eqs.(\ref{Eq:omegah2inflatonsmallmj}) and (\ref{Eq:omegah2inflatonapprox1}).
It is easy to understand, as in the
case of inflaton reheating, lower values of $M_{\rm in}$ produces
less dark matter particles, see Eq.(\ref{Eq:ntotmjless}), and then necessitates higher dark matter masses. The situation is the 
opposite for the intermediate value of
$M_{\rm in}$ when PBH dominates the reheating because of the dilution effect described by 
Eq.(\ref{Eq:nt3}).

%$\xi=3.8$ and $g_*(T_{\rm BH})=106.5$. $n_{\rm S}^{\rm BH}$ is the number density of the DM arising from the PBH evaporation. Here, $\Gamma_{{\rm BH}\rightarrow{S}}$ is the non-thermal production term for DM (originating from PBH evaporation) and can be written as 

\section{Refinements} \label{refinement}

\subsection{The case for extended mass distribution}

Depending on the underlying mechanism that governs their formation, PBHs may exhibit extended mass distribution that is contingent on the power spectrum of primordial density perturbations and the equation of state of the Universe at the time of their formation (see Ref.~\cite{Carr:2017jsz}), for example power-law~\cite{Carr:1975qj}, log-normal \cite{Dolgov:1992pu,Green:2016xgy,Dolgov:2008wu}, critical collapse \cite{Carr:2016hva,Musco:2012au,Niemeyer:1999ak,Yokoyama:1998xd}, or metric preheating \cite{Martin:2019nuw,Martin:2020fgl,Auclair:2020csm}, among others.
In the current section, we consider the class PBHs with power-law shape mass function of the form:
\beqa
f_{\rm PBH}(M_i,t_i) = \begin{cases}
    C M_i^{-\alpha}, & {\rm for}\, \,  M_{\rm min} \leq M_i \leq M_{\rm max} \, \\
    0, & {\rm otherwise}\, ,
\end{cases}
\label{Eq:power-law-dist}
\eeqa
where $t_i$ and coefficient $C$ are respectively, the initial time and the overall normalization factor. $M_{\rm min}$ and $M_{\rm max}$ represent the minimum and the maximum PBH masses, respectively. We parameterize the width of PBHs masses range by two parameters $M_{\rm in}$ and $\sigma$, such that $M_{\rm min} = M_{\rm in} 10^{-\sigma}$ and $M_{\rm max} = M_{\rm in}$. The parameter $\alpha=\frac{2+4\omega_\phi}{1+\omega_\phi}$ (see Ref. \cite{RiajulHaque:2023cqe,Cheek:2022mmy}).

Then the evolution equations for ${\rho}_{\rm BH}$ and ${\rho}_R$ become:
%\beqa 
\begin{multline}
\frac{d{\rho}_R}{da}+4 \frac{\rho_R}{a} = \frac{\Gamma_\phi\,\rho_\phi\,(1+w_\phi)}{a\,H}  \\ -  \frac{a^3}{a^3_{\rm in}} \int_{\widetilde M}^{\infty} \frac{dM}{da} f_{\rm PBH}(M_i, t_i) dM_i 
%\label{Eq:evolution-eq-rad-ext}
\\
\frac{d{\rho}_\text{BH}}{da}+3\frac{\rho_{\rm BH}}{a} =  \frac{a^3}{a^3_{\rm in}} \int_{\widetilde M}^{\infty} \frac{dM}{da} f_{\rm PBH}(M_i, t_i) dM_i \,
\label{Eq:evolution-eq-pbh-ext} 
\end{multline}
%\eeqa 
where the lower bound $\widetilde M$ allows to ensure that at time $t$ only the non-evaporated PBHs with mass $M_i$ larger than $\widetilde M$ contribute to the energy density, and is given by:
\beqa 
\widetilde M(a) = \left(\frac{2\sqrt{3} \epsilon}{1+\omega_\phi} \right)^{1/3} \left(\frac{M_P^5}{\sqrt{\rho_{\rm end}}} \right)^{1/3} \left(\frac{a}{a_{\rm in}} \right)^{\frac{1}{2}(1+\omega_\phi)} \, .
\label{eq:eva_ini_mass}
\eeqa

In this scenario, we employed a modified version of the package called \hyperlink{https://github.com/yfperezg/frisbhee}{\tt FRISBHEE} \cite{Cheek:2021odj,Cheek:2021cfe,Cheek:2022dbx,Cheek:2022mmy}. This modified version incorporated the inflaton into the evolving system, enabling us to solve a set of evolution equations and calculate the relic abundance. This approach is necessary due to the intricacies introduced by the presence of integrals on the right-hand side of Eq.(\ref{Eq:evolution-eq-pbh-ext}), which makes the situation somewhat more complex than the monochromatic scenario.

Before presenting the results, let's recall that the reheating through PBHs, after a regime of PBH domination, happens when they completely evaporate. Hence, as shown in \cite{RiajulHaque:2023cqe}, the choice of the mass function that extends to lower values, with the maximal initial mass $M_{\rm max}$ corresponding to the monochromatic mass $M_{\rm in}$, guarantees that the complete evaporation of PBHs in both cases is achieved at the same epoch. Therefore, the reheating temperature would be somewhat the same in both cases for a given $M_{\rm in}$. 

\begin{figure}[!ht]
 %\centering
	\includegraphics[width=\columnwidth,height=5.47cm]   {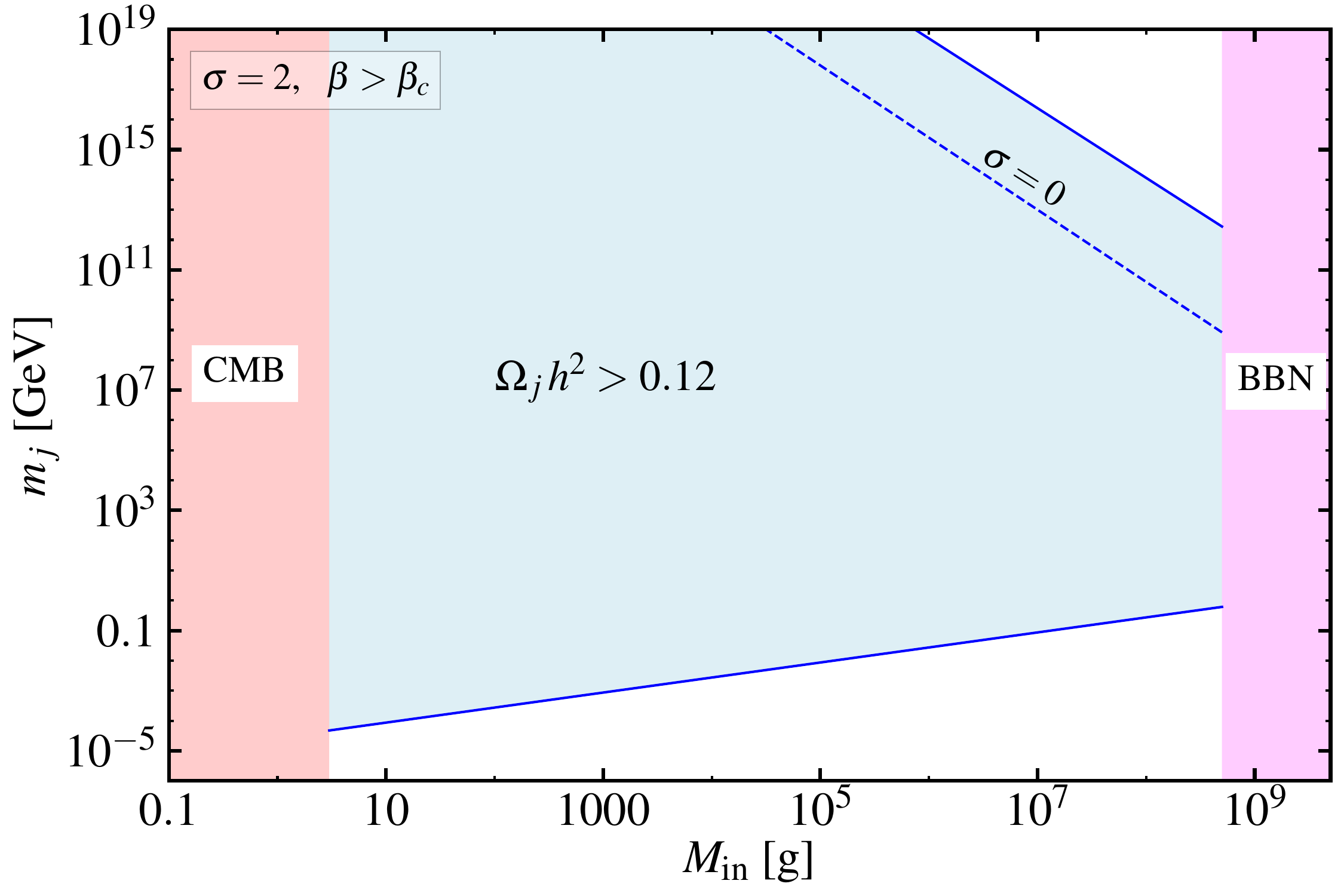}
     \caption{\it Allowed region in the ($M_{\rm in}\,,m_j$) plane for $\beta >\beta_c$ in the case of the power-law extended mass distribution, with $\sigma=2$ (see text for details). The dashed blue line corresponds to the monochromatic ($\sigma=0$) scenario. BBN and CMB bounds exclude the magenta and red-shaded regions, respectively.}
	\label{n=4_mDM_vs_MBH_beta>betaBH}
\end{figure}

Our findings yield similar results for both the extended and monochromatic PBH mass spectra, as depicted in Fig.(\ref{n=4_mDM_vs_MBH_beta>betaBH}), under the conditions of $\beta > \beta_c$ and $\sigma = 2$, which corresponds to the range $M_{\rm in} \in [10^{6}, \, 10^{8}]\,$g. Nevertheless, certain distinctions are noteworthy.

\noindent
We observe that in the light DM mass region, $m_j<T_{\rm BH}^{\rm in}$, there is no substantial difference between the monochromatic and extended cases, as can be seen in the lower line of Fig.(\ref{n=4_mDM_vs_MBH_beta>betaBH}). This happens because the number of DM particles produced is $\propto (M_{\rm in})^2$, and so the dominant contribution comes from
heaviest PBH, which in our case corresponds to monochromatic mass.

\noindent

However, in the high DM mass region, $m_j>T_{\rm BH}^{\rm in}$, there is a substantial difference. First, the number density of DM is larger compared with the monochromatic case because not only the largest mass $M_{\rm in}$ produces DM particles, but as soon as the lighter population of PBHs start to evaporate, they will produce DM particles $\propto (m_j)^{-2}$ as well. Consequently, a much wider parameter space of $m_j$ is excluded. The exclusion region for a given $M_{\rm in}$ would be similar to that of the monochromatic scenario when the PBH initial mass $\gtrsim M_{\rm in} 10^{-\sigma}$. This can be observed in Fig.(\ref{n=4_mDM_vs_MBH_beta>betaBH}) where, for example, for $M_{\rm in} \sim 10^8 $g  and $\sigma = 2$, avoiding overproduction of DM requires $m_j \gtrsim 10^{14}$ GeV, corresponding to roughly to the excluded region for the monochromatic case ($\sigma = 0$) when $M_{\rm in} \sim 4 \times 10^6 $g as can be deduced from the upper dashed blue line.

In light of these results, some comments are in order:
The exclusion region for $m_j$ is strongly dependent on the size of the width of the distribution; in such a way the larger the $\sigma$, the stronger the constrain on $m_j$, that is, the larger the mass of DM particles is necessary not to overproduce it.
We also note that if the distribution extends to larger masses such that $M_{\rm in} = M_{\rm min}$, the reheating temperature and the DM number density can be very affected depending on the width of the mass function, since the larger masses than $M_{\rm in}$ would have larger lifetime compared to the monochromatic scenario. This will affect both the low and high DM mass regimes.

\subsection{Limit on the DM mass from warm dark matter (WDM) constraints}
\begin{figure}
    \centering
    \includegraphics[scale=0.42]{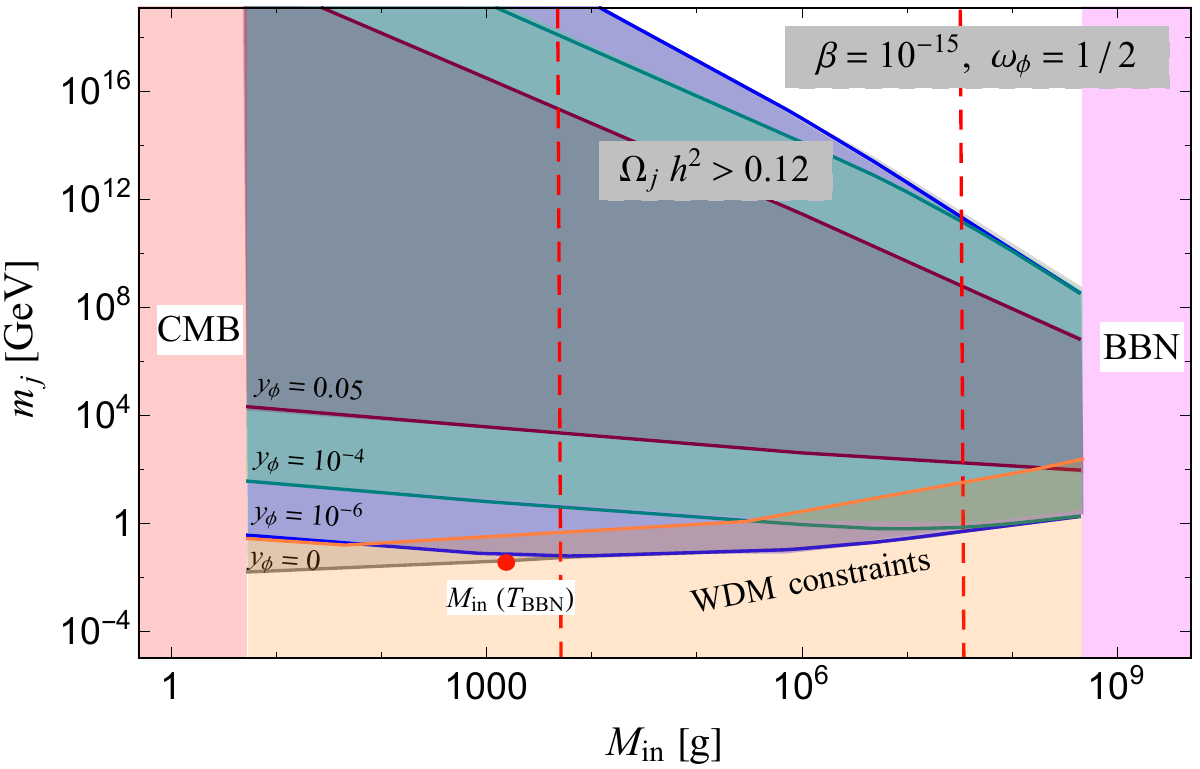}~~ \caption{\it
    %Allowed region in the $(M_{\rm in},\,m_j)$ plane for a fixed value of $\beta=10^{-15}$ and $w_\phi=\frac{1}{2}$ together 
    Same as Fig.(\ref{Fig:mdmvsmbh2}) taking into account the restriction from the warm dark matter constraint
    (orange shaded region). %Different regions are for different coupling values. 
%    The orange-shaded region is restricted from the WDM constraints. Other details of this plot are the same as Fig.(\ref{Fig:mdmvsmbh2}). 
}
    \label{fig:wdm}
\end{figure}

\subsubsection{Generalities}

The DM from PBH evaporation has a large 
initial momentum. Higher initial momentum indicates a large free 
streaming length, which might erase 
small-scale structures. 
 Indeed, if boosted at production time, the classical limit on warm dark matter ($m_j \gtrsim 3$ keV) coming from structure formation or Lyman-$\alpha$ constraints needs to be revisited. The idea is simple. A straightforward calculation shows that the free streaming length $\lambda_{FS}$ can be approximated by \cite{mybook}
\beq
\lambda_{FS}\simeq 70~{\rm Mpc}\frac{1~\rm eV}{T_{nr}}\,,
\label{Eq:freestreaming}
\eeq
where $T_{nr}$ is the temperature 
at the time when the DM becomes non-relativistic, or $p \lesssim m_j$, $p$ is the momentum of the DM. If $p\sim T$
at production time, because $T$ and 
$p$ redshifts as $a^{-1}$, the condition can indeed be read $T_{nr}\sim m_j$, and the classical limits apply. This is the typical case for WIMP or FIMP candidates.
However, if the dark matter momentum $p$ is boosted by a process at production time, $p= \gamma T$,
with $\gamma \gg 1$
the condition $p\sim m_j$ becomes $T_{nr}\sim \frac{m_j}{\gamma}$, and Eq.(\ref{Eq:freestreaming}) becomes
\beq
\lambda_{FS}\simeq 70~{\rm Mpc}\frac{1~\rm eV}{m_j}\gamma\,,
\eeq
transforming the condition from $m_j\gtrsim 3$ keV to $m_j \gtrsim 3~ \gamma$ keV. This is exactly what is happening in the case of production from PBH evaporation because the dark matter momentum at evaporation 
$p_{\rm ev}\sim T_{\rm BH}^{\rm in} \gg \trh$.
We propose to study in more detail each cases analyzed previously in this context.

\subsubsection{$\beta>\beta_{\rm c}$}

Since DM 
particles have no interaction with 
other particles from the evaporation 
point to the present day
%, their 
%momentum simply redshifts, and 
we have momentum value at the present day

\begin{equation}\label{momentum-presentday}
p_{\rm 0}=\frac{a_{\rm ev}}{a_{\rm0}}\,p_{\rm ev}=\frac{a_{\rm ev}}{a_{\rm eq}}\,\frac{\Omega_{\rm R}}{\Omega_{\rm m}}\,p_{\rm ev}\,
\end{equation}
where at present-day radiation relic 
abundance $\Omega_{\rm R}=5.4\times 10^{-5}$ and matter relic abundance $\Omega_{\rm m}\simeq 0.315$. $a_{\rm eq}$ and $a_{\rm ev}$ are the scale 
factor at the radiation-matter 
equality and evaporation point, respectively. 
%Now let us calculate $p_{\rm 0}$ for different scenarios.
In the limit, $\beta>\beta_{\rm c}$, 
PBH evaporation completes during PBH 
domination, and comoving entropy density is conserved from the end of 
the evaporation point to the present 
day. In this context, $T_{\rm ev}=T_{\rm RH}$ and using entropy 
conservation, $p_{\rm 0}$ can be rewritten as
\begin{equation}\label{Eq:ppbh}
p_{0}=\left[\frac{g_{\rm eq}^s}{g_{\rm RH}^s}\right]^{1/3}\frac{T_{\rm eq}}{T_{\rm RH}}\frac{\Omega_R}{\Omega_m}\,p_{\rm ev}\,, 
\end{equation}
where $g_{\rm RH}^s$ and $g_{\rm eq}^s$ represent the 
effective degrees of freedom for entropy at the end of reheating and radiation-matter equality, respectively.  
$T_{\rm eq}=0.8$ eV is the radiation temperature at 
radiation-matter equality. In the case of light DM, $m_j\ll T_{\rm BH}^{\rm in}$, the average momentum of the light DM matter particle radiated by a 
PBH, $p_{\rm ev}\sim T_{\rm BH}^{\rm in}$. Moreover, the usual 
velocity of the warm DM at present, which decoupled 
while they are relativistic, is assumed to be \cite{Bode:2000gq}

\bea \label{Eq:velcons}
v_{\rm WDM}\simeq 3.9\times 10^{-8}\,\left(\frac{\rm keV}{m_{\rm WDM}}\right)^{\frac{4}{3}}\,.
\eea
Several experiments put constraints on 
the WDM, such as HIRES/MIKE Lyman-$\alpha$ forest data sets and XQ-$100$, the MCMC analysis  restricts WDM mass $m_{\rm WDM}> 5.3$ keV at $2\,\sigma$ range \cite{Irsic:2017ixq}. In references \cite{Viel:2013fqw,Hoof:2018hyn}, 
using HIRES/MIKE, the authors 
obtained the bound on WDM $m_{\rm WDM}>3.3$ keV and $>3.95$ keV 
using SDSSIII/BOSS. Throughout our 
analysis, we choose the 
restriction on the mass of the WDM, $m_{\rm WDM}>3.3$ keV. Now 
utilizing the above equations (\ref{Eq:ppbh}) and (\ref{Eq:velcons}) together with (\ref{Eq:trh1}), one can find
\beq \label{wdmpbh}
\frac{m_j}{\rm GeV}\geq 7\times10^{-7}\left(\frac{m_{\rm WDM}}{\rm keV}\right)^{\frac{4}{3}}\left(\frac{M_{\rm in}}{M_P}\right)^{\frac{1}{2}} \,,
\end{equation}
where $m_{\rm WDM}\sim 3.3$ keV. For example, if one takes 10 g of initial PBH mass, the DM mass bound turns out to be $m_j > 5.2 \times 10^{-3}$ GeV. Note that this bound does not depend on the inflaton EoS and PBH fraction as expected.
We also recover the naive constraint we obtained previously remarking that the boost factor $\gamma=\frac{p_{\rm ev}}{\trh} \sim \frac{T_{\rm BH}^{\rm in}}{\trh} \sim \sqrt{\frac{\Min}{M_P}}$. 
%%%%%%%%%%%%%%%%%%%%%%%%%%%%%%%%%%%%%%%%%%%%
\subsubsection{$\beta<\beta_{\rm c}$ (PBH reheating)}
In the limit $\beta<\beta _{\rm c}$, if the coupling value $y_\phi<y_\phi^c$, PBH decay determines reheating temperature. Thus, after the evaporation point to the present day, the comoving entropy energy density is conserved, and the expression for $p_0$
\begin{equation}\label{Eq:pbhrehe}
p_{0}=\left[\frac{g_{\rm eq}^s}{g_{\rm ev}^s}\right]^{1/3}\frac{T_{\rm eq}}{T (\aev)}\frac{\Omega_R}{\Omega_m}\,\frac{M_P^2}{\Min}\,, 
\end{equation}
where $g_{\rm ev}^s$ represents the 
effective degrees of freedom for entropy at the end of evaporation. In this scenario, PBH evaporates during inflaton domination, and evaporation temperature can be calculated from $\rbh (\aev)$ (see, for instance, Eq.(\ref{Eq:nbhaev}))
\beq \label{Eq:Tev}
T(\aev)\simeq \left(\frac{\rbh}{\alpha_T}\right)^{\frac{1}{4}}=\left(\frac{48\pi^2\beta}{\alpha_T}\right)^{\frac{1}{4}}\tilde{\mu_1}M_P\left(\frac{M_P}{\Min}\right)^{\frac{3+w_\phi}{2\,(1+w_\phi)}}\,,
\eeq
where $\tilde{\mu_1}=\left(\frac{\gamma^{w_\phi}\,\epsilon}{2\pi(1+w_\phi)}\right)^{\frac{1}{2\,(1+w_\phi)}}$. Utilizing Eq.(\ref{Eq:pbhrehe}) and (\ref{Eq:Tev}), the restriction on the DM mass 
\beq
\frac{m_j}{\rm GeV}\geq \frac{6.1\times10^{-7}}{\tilde{\mu_1}\,\beta^{\frac{1}{4}}}\left(\frac{m_{\rm WDM}}{\rm keV}\right)^{\frac{4}{3}}\left(\frac{M_{\rm in}}{M_P}\right)^{\frac{1-w_\phi}{2\,(1+w_\phi)}} \,,
\eeq
For example, if one takes 10 g of initial PBH, the DM mass bound turns out to be $m_j > \frac{5.1 \times 10^{-5}}{\beta^{\frac{1}{4}}}$ GeV, for $w_{\phi} =1/2$. 
 Now let us move our discussion to the case of inflation reheating.
 %%%%%%%%%%%%%%%%%%%%%%%%%%%%%%%%%%%%%%%%%%%%%%%%%
\subsubsection{$\beta<\beta _{\rm c}$ (Inflaton reheating)}
If the coupling strength $y_\phi>y_\phi^{\rm c}$ and $\beta<\beta_{\rm c}$, inflaton coupling determines reheating temperature. In addition to that depending on how strong the coupling is $y_\phi>y_\phi^{\rm th}>y_\phi^{\rm c}$ or $y_\phi^{\rm c}<y_\phi<y_\phi^{\rm th}$, reheating happen before and after the evaporation point, respectively. The restriction would be different in these two cases.
\begin{itemize}
    \item {$\aev<\arh$ :} Once $y_\phi^{\rm c}<y_\phi<y_\phi^{\rm th}$, reheating occurs after completion of the evaporation process. One can find the ratio $\aev/a_{\rm eq}$ as
\beq \label{Eq:aevaeq1}
\frac{\aev}{a_{\rm eq}}=\frac{\aev}{\arh}\frac{\arh}{a_{\rm eq}} =\left(\frac{g_{\rm eq}^s}{g_{\rm RH}^s}\right)^{1/3}\frac{T_{\rm eq}}{\trh}\tilde{\mu_2}\left(\frac{\trh^2}{M_P^2}\frac{\Min^3}{M_P^3}\right)^{\frac{2}{3\,(1+w_\phi)}}\,,
 \eeq
where $\tilde{\mu_2}=\left(\frac{1+w_\phi}{2\,\epsilon}\sqrt{\frac{\alpha_T}{3}}\right)^{\frac{2}{3\,(1+w_\phi)}}$. To derive above equation we use Eq.(\ref{Eq:aevarh1}). Now, upon substitution of the above Eq.(\ref{Eq:aevaeq1}) into (\ref{Eq:ppbh}) and employing Eq.(\ref{Eq:velcons}), we get 
\bea
&&
\frac{m_j}{\rm GeV}\geq 1.2\times 10^{-6}\,\tilde{\mu_2}\left(\frac{m_{\rm WDM}}{\rm keV}\right)^{\frac{4}{3}}
\nonumber
\\
&&
\left(\frac{\trh}{M_P}\right)^{\frac{1-3\,w_\phi}{3\,(1+w_\phi)}}\left(\frac{\Min}{M_P}\right)^{\frac{1-w_\phi}{1+w_\phi}}\,.
\eea
For example, if one takes 10 g of initial PBH mass, the DM mass bound turns out to be $m_j > \left(\frac{\trh}{\rm GeV}\right)^{-1/9} 6.6 \times 10^{-2}$ GeV, for $w_{\phi} =1/2$.

\item {$\aev>\arh$ :} For strong coupling $y_\phi>y_\phi^{\rm th}>y_\phi^{\rm c}$, the reheating process completes even before the evaporation point, and the leading decay process takes place in a radiation-dominated background. In this case, $T(\aev)$ can be calculated from the Hubble parameter at the evaporation point, which is related to $\Gamma_{\rm BH}$, $H(\aev)=\frac{\Gamma_{\rm BH}}{2}=\frac{3\,\epsilon}{2}\frac{M_P^4}{\Min^3}$. The evaporation temperature
\beq \label{Eq:Tevrad}
T(\aev)=\left(\frac{3\,M_P^2\,H(\aev)^2}{\alpha_T}\right)^{\frac{1}{4}}=\left(\frac{27\,\epsilon^2}{4\,\alpha_T}\right)^{\frac{1}{4}}\frac{M_P^{\frac{5}{2}}}{\Min^{\frac{3}{2}}}\,.
\eeq
Since in this scenario, after evaporation no entropy injection in the Universe, combining Eq.(\ref{Eq:pbhrehe}) and (\ref{Eq:Tevrad}), one can find
\beq
\frac{m_j}{\rm GeV}\geq 8.1\times10^{-7}\left(\frac{m_{\rm WDM}}{\rm keV}\right)^{\frac{4}{3}}\left(\frac{M_{\rm in}}{M_P}\right)^{\frac{1}{2}} \,.
\eeq
\end{itemize}
This particular constraint is similar to the one previously discussed for the PBH-dominated case (see Eq.(\ref{wdmpbh})).
This behavior was expected because it corresponds to our naive estimate
of the boost factor $\gamma$ discussed previously.

We depicted in Fig.(\ref{fig:wdm}) how DM parameter space presented in Fig.(\ref{Fig:mdmvsmbh2}) is modified if one considers WDM constraints, which we show in orange shaded region. One important outcome of Fig.(\ref{fig:wdm}) is that for $m_j<T_{\rm BH}^{\rm in}$, the allowed DM masses in the context of purely PBH reheating is severely restricted due to the violation of the WDM limit.

\subsection{PBH evaporation: Comparison with the exact greybody factor}
PBH mass reduction rate is crucially dependent on evaporation process, produced particles' spin, and the angular momentum of the BHs \cite{Cheek:2021odj}. Throughout our 
analysis we dealt with the 
Schwarzschild BHs, 
%so BHs have zero angular momentum and leave the analysis for the spinning BHs for a 
%future study. 
To be precise the rate of change of BH mass is calculated upon integrating over the phase space and summing over different species as
\beq
\frac{dM_{\rm BH}}{dt}=-\sum_j\int_{0}^{\infty}E_j\,\frac{\partial^2N_j}{\partial p\,\partial t}\,dp=-\epsilon(M_{\rm BH})\,\frac{M_P^4}{M_{\rm BH}^2}\,,
\eeq
where $\frac{\partial^2N_j}{\partial p\,\partial t}$ represents the emission rate of any species $j$ of 
mass $m_j$ and spin $s_j$ with 
degrees of freedom $g_j$ in time interval $dt$ and momentum lies 
within\footnote{For details calculation, see 
Ref.\cite{Cheek:2021odj}.} $(p,\,p+dp)$  and $E_j=\sqrt{m_j^2+p^2}$. The BH mass-dependent evaporation function 
$\epsilon(M_{\rm BH})$ can be expressed as
\beq
\epsilon (M_{\rm BH})=\sum g_j\,\epsilon_j (z_j)\,,
\eeq
where 
\beq
\epsilon_j(z_j)=\frac{27}{128\,\pi^3}\,\int_{0}^{\infty} \frac{\Psi_{\rm s_j}\left(x_j^2-z_j^2\right)}{\exp(x_j)-(-1)^{2\,s_j}}\,x_j\,dx_j\,,
\eeq
where $x_j$ and $z_j$ are the dimensionless parameter defined as $x_j=E_j/T_{\rm BH}$, $z_j=m_j/T_{\rm BH}$. And $\Psi_{\rm s_j}$ is the reduced greybody factor defined as the ratio between the exact greybody factor to its value in the {\it geometrical-optics} limit
\beq
\Psi_{\rm s_j} (x)=\frac{\sigma_{\rm sj}}{\sigma_{\rm sj}|_{\rm go}}\,.
\eeq
In the {\it geometrical-optics} limit, the greybody factor assumes, $\sigma_{\rm sj}|_{\rm go}=\frac{27}{64\pi}\,\frac{M_{\rm BH}^2}{M_P^4}$ \cite{Page:1976df,Page:1977um,MacGibbon:1990zk,MacGibbon:1991tj} and the evaporation function for massless particle turns out as
\beq
\epsilon_j(0)=\frac{27}{4}\frac{\xi\,\pi\,g_j}{480}\,,
\eeq
where $\xi=(1,7/8)$ for bosons and fermions respectively, which is the limit we took. Consequently, PBH mass evolution follows Eq.(\ref{Eq:mbh}).
%\beq
%\frac{dM_{\rm BH}}{dt}=\frac{27}{4}\frac{\pi\,g_*(T_{\rm BH})}{480}\,.
%\eeq
%where $g_*(T_{\rm BH})\sim 106.75$ counting for the Standard Model particles degrees of freedom at $T_{\rm BH}$.

One important point to note is that the evaporation function in the {\it geometrical-optics} 
limit nearly matches with the actual evaporation function for scalar and fermionic particles, however for fermion mass $m_j>4\, T_{\rm BH}$ our calculation is slightly underestimated compare to the actual one, as we can see from the Fig.(2) of 
Ref.\cite{Cheek:2021odj} where they plotted the evaporation function $\epsilon_j(z_j)$ as function of $z_j$. Finally, we can add 
that for scalar (spin zero) and 
fermionic (spin half) particles, the {\it geometrical-optical} limit works 
fine,  whereas for higher spin 
particles such as spin-1 and spin-2 
particles, we would need to take the exact spectrum for the accurate analysis.

\section{Conclusion}\label{conclusion}

In this study, we have compared in detail the DM parameter space in the background of the reheating phase dynamically obtained from 
two chief systems in the early Universe: the inflaton $\phi$ and the primordial black 
holes. The DM is assumed to be produced purely gravitationally from the PBH decay, 
not interacting with the thermal bath and the inflaton. Within this context, 
The population of the primordial black holes behaves like dust, whereas the behavior 
of inflaton depends strongly on its equation of state after the inflationary phase, which 
in turn depends on the exponent of the potential $V(\phi)\propto \phi^n$, 
$w_\phi=\frac{n-2}{n+2}$. Depending upon 
the dynamics of reheating, we showed that a 
large range of initial PBH masses $M_{\rm in}$ and fraction $\beta$ can lead to the right amount of relic abundance.

If PBHs dominate the background dynamics ($\beta>\beta_{\rm c}$), the reheating process becomes 
insensitive to the inflaton and the PBH 
fraction $\beta$. Therefore, it is the PBH mass $M_{\rm in}$ that solely controls the 
DM abundance as well as the reheating temperature $\trh$. In this scenario, if $m_j < T_{\rm BH}^{\rm in}$, 
with increasing DM mass, we need to increase the value of $M_{\rm in}$ to increase the dilution such that the DM 
number density decreases to avoid the DM 
overproduction i.e., $m_j\propto \sqrt{M_{\rm in}}$. Another possibility is to have DM mass $m_{j} > T_{\rm BH}^{\rm in}$, where the Boltzmann suppression 
naturally reduces the DM production and 
hence opens up the lower $M_{\rm in}$ values for which abundance could be satisfied. This 
scenario is nicely illustrated by
Fig.(\ref{Fig:mdmvsmbh}). 

If one considers $y_\phi<y_\phi^{\rm c}$ and $\beta<\beta_{\rm c}$ that allows
PBH radiation to govern the reheating while the inflaton dominates the energy budget during the whole process. For such case, the allowed DM region can be extended, and that is solely dependent on the inflaton equation of state $w_\phi$. As an example we have shown the results for $w_\phi={1}/{2}$  in Fig.(\ref{fig:dmbetalbetac}).
The main conclusion of our analysis is that two 
very secluded region of PBH parameters can, at the same time
ensures a successful reheating while still producing the right amount of dark matter with mass $m_j$: 
$10^{-4} ~\mbox{GeV} \lesssim m_j \lesssim 1$ GeV (corresponding to $ m_j \ll T^{\rm in}_{\rm BH}$), and $10^{19} ~\mbox{GeV} \gtrsim m_j \gtrsim 10^8$ GeV (corresponding to $ m_j \gg T^{\rm in}_{\rm BH}$).

In the case of extended mass function with power-law distribution extending to lower PBH mass values, the limits in the lower DM regime ($m_j < T_{\rm BH}^{\rm in}$) remain unchanged compared to the monochromatic scenario. However, in the high DM mass regime ($m_j > T_{\rm BH}^{\rm in}$) gets modified depending on the width of the distribution. For instance, with $\sigma = 2$, the constraint on the range of allowed values for $m_j$ becomes $10^{19}\, \text{GeV} \gtrsim m_j \gtrsim 10^{12} \, \text{GeV}$.

If the energy budget {\it and} the reheating is dominated by the inflaton, the range of allowed DM mass is widened and depends 
strongly on the Yukawa coupling of the inflaton to the Standard Model, $y_\phi$ as one can see in
Fig.(\ref{Fig:mdmvsmbh2})
which can be considered as the master plot of our work.
In comparison with the PBH reheating, a noticeable difference in DM parameter space can be observed in both $m_j < T^{\rm in}_{\rm BH}$ and $m_j > T^{\rm in}_{\rm BH}$ case.  Particularly 
when $m_j > T^{\rm in}_{\rm BH}$, for which DM yield is large for lower PBH and DM mass, $\Omega_j h^2\propto M_{\rm in}^{-5/3} m_j^{-1}$ compare to PBH reheating  $\Omega_j h^2\propto M_{\rm in}^{-13/6} m_j^{-1}$ for $w_{\phi}=1/2$. In 
this case, additional entropy injection from inflaton dilutes the yields. Consequently, it allows lower PBH 
masses to not overclose the Universe.
Hence, the increment of the inflaton coupling $y_{\phi}$ widen the mass 
range by rendering both lower values of $m_j$ and $M_{\rm in}$ viable for $y_\phi\sim 0.05$ as we see 
in Fig.(\ref{Fig:mdmvsmbh2}). 
Interestingly for $m_j < T^{\rm in}_{\rm BH}$ case, on the other hand, the dilution due to entropy 
injection effects oppositely on the DM yields, $\Omega_j h^2\propto M_{\rm in}^{1/3} m_j$ for $w_{\phi}=1/2$, 
rendering it under abundant. Thus, one must increase 
the $m_j$ value in inflation reheating to obtain the correct DM yield with increasing $y_{\phi}$.
Indeed, the decoupling between the 
reheating process (completed by the inflaton) and the dark matter production (generated by the PBHs) allows for a larger range of dilution factor through the injection of the entropy from inflaton decay. As a consequence, larger DM mass
are necessary for the same amount of relic abundance.
Moreover, in PBH reheating, for $m_j < T^{\rm in}_{\rm BH}$, $m_j\propto M_{\rm in}^{1/6}$, whereas inflation reheating suggests $m_j\propto M_{\rm in}^{-1/3}$, which is a completely opposite behavior which we can see in Fig.(\ref{Fig:mdmvsmbh2}).

We have also included the warm dark matter constraints from structure formation and Lyman-$\alpha$ forest. Indeed, the dark matter momentum at evaporation time being
$p_{\rm ev}\sim T_{\rm BH}^{\rm in}\gg \trh$, the typical limit $m_j \gtrsim 3$ keV needs
to be revisited. We considered this boost factor in our analysis, which makes the dark matter candidate relativistic for longer. We found that the 
region $m_j < T_{\rm BH}^{\rm in}$ previously allowed for pure PBH reheating is now
excluded due to warm dark matter limit, whereas the region $m_j>T_{\rm BH}^{\rm in}$ stays unaffected as 
we can see in Fig.(\ref{fig:wdm}). On the other 
hand, the presence of the inflaton produces a sufficiently large amount of entropy, decreasing the free streaming length significantly.
In this case, the warm dark matter constraint does not affect our result either, as we also see in Fig.(\ref{fig:wdm}).

In conclusion, we see that the combination of two chief systems of the world, even with very different phenomenology and dynamics, can considerably enlarge the parameter space allowed by the cross-constraints from reheating and the relic abundance.

\vspace{0.5cm} 
\noindent
\acknowledgements
E.K. and Y.M. want to thank L. Heurtier for extremely valuable discussions during the completion of our work. This project has received support from the European Union's Horizon 2020 research and innovation programme under the Marie Sklodowska-Curie grant agreement No 860881-HIDDeN, and the IN2P3 Master Projet UCMN. 
M.R.H wishes to acknowledge support from the Science and Engineering Research Board (SERB), Government
of India (GoI), for the SERB National Post-Doctoral fellowship, File Number: PDF/2022/002988. DM wishes
to acknowledge support from the Science and Engineering Research Board (SERB), Department of Science and
Technology (DST), Government of India (GoI), through the Core Research Grant CRG/2020/003664.  DM also thanks the Gravity and High Energy
Physics groups at IIT Guwahati for illuminating discussions.
The work of E.K. was supported by the grant "Margarita Salas" for the training of young doctors (CA1/RSUE/2021-00899), co-financed by the Ministry of Universities, the Recovery, Transformation and Resilience Plan, and the Autonomous University of Madrid.

\appendix

\section{Expression for the critical coupling
$y_\phi^c$}
\label{appendixA}
In the standard reheating scenario,  the reheating process is not instantaneous, and during this phase, the inflaton energy density transfers to the daughter particles, mostly to the SM particles, setting proper initial conditions for the Big Bang Nucleosynthesis (BBN). In principle, considering different gravitational and non-gravitational couplings, there are several possibilities for reheating. However, in this analysis, we are only interested in the fermionic coupling with interaction Lagrangian $y_\phi\phi\bar{f}f$. Taking such a scenario, we have the radiation energy density
\bea
\rho_R^{\rm D}(a) &=& \frac{y_\phi^2}{8 \pi}\, \lambda^{\frac{1-w_\phi}{2\,(1+\,w_\phi)}}  \alpha_n\,M_P^4
\left(\frac{\rhoe}{M_P^4}\right)^{\frac{3}{2}-\frac{1}{1+w_\phi}} \left(\frac{a}{\ae}\right)^{-4}
\nonumber
\label{Eq:rhorwobh}
\\
&&
\times
\left[\left(\frac{a}{\ae}\right)^{\frac{5-9\,w_\phi}{2}}-1\right] \,,
\eea
where $a_{\rm end }$ is the scale factor associated with the end of inflation, and $\lambda$ is related to the mass scale $\Lambda$ of the $\alpha-$ attractor potential \cite{Ellis:2013nxa,Kallosh:2013yoa},  $\lambda=\left(\frac{\Lambda}{M_P}\right)^4\,\left(\frac{2}{3\,\alpha}\right)^{\frac{n}{2}}$, potential which has the form

    \begin{equation} \label{a}
V(\phi) = \Lambda^4 \left[  1 - e^{ -\sqrt{\frac{2}{3\,\alpha}}\dfrac{\phi}{M_P} } \right]^{n}\,.
\end{equation}
The parameter $\lambda$ can be represented  in terms of the CMB observables, such as the amplitude of the inflaton fluctuation $A _{\mathcal{R}}$ and scalar spectral index $n _s$ as \cite{Drewes:2017fmn}
\beqa
 \lambda = & &\left(\frac{2}{3\,\alpha}\right)^{\frac{n}{2}} {\left(\frac{3\pi^2 r A_{\mathcal{R}}}{2}\right)^4 }  \nonumber \\
 & & \times \left[\frac{n^2+n+\sqrt{n^2+3\alpha(2+n)(1-n_s)}}{n(2+n)}\right]^{n}
 \eeqa

The above equation suggests that the evolution of the radiation energy density is different for $w_\phi>5/9$ ($n>7$) and $w_\phi<5/9$ ($n<7$). As a consequence, the end of the reheating, which is defined at the point of $\arh$ where $\rho_\phi(\arh)=\rho_{\rm R} (\arh)=\rho_{\rm RH}$ would be different for $n>7$ and $n<7$.
In the case of $n<7$, $\arh$ can be written as \cite{Garcia:2020wiy,Garcia:2020eof},
\bea \label{arhnl7}
\frac{\arh}{a_{\rm end}}=\left[\frac{y_\phi^2}{8\pi}\alpha_n\left(\frac{\lambda\,M_P^4}{\rho_{\rm end}}\right)^{\frac{1-w_\phi}{2\,(1+w_\phi)}}\right]^{\frac{2}{3\,(w_\phi-1)}}\,.
\eea
However for $n>7$ one can find
\bea \label{arhng7}
\frac{\arh}{a_{\rm end}}=\left[-\frac{y_\phi^2}{8\pi}\alpha_n\left(\frac{\lambda\,M_P^4}{\rho_{\rm end}}\right)^{\frac{1-w_\phi}{2\,(1+w_\phi)}}\right]^{\frac{1}{1-3\,w_\phi}}\,.
\eea

Upon substitution, the expression for $\arh$ Eq.(\ref{arhng7}) and (\ref{arhng7}) into (\ref{Eq:rhorwobh}) $\rho_{\rm RH}$ can be written as,
\bea \label{Eq:rhorh1}
\rho_{\rm RH}^{\rm D}=\left(\frac{y_\phi^2}{8\,\pi}\,\alpha_n\right)^{\frac{2\,(1+w_\phi)}{1-w_\phi}}\lambda\,M_P^4\,.
\eea
Whereas, for $n>7$,
\beq \label{Eq:rhorh2}
\rho_{\rm RH}^{\rm D}=\left(\frac{y_\phi^2}{8\,\pi}\,\alpha_n\right)^{\frac{3\,(1+w_\phi)}{3\,w_\phi-1}}\left(\lambda\,M_P^4\right)^{\frac{3\,(1-w_\phi)}{2\,(3\,w_\phi-1)}}\,\rho_{\rm end}^{\frac{5-9\,w_\phi}{2\,(1-3\,w_\phi)}}.
\eeq
The expression for the critical coupling $y_\phi^c$ below which value PBH-driven reheating happens should be followed
%%%%%%%%%%%%%%%%%%%%%%%%%%%%%%%%%%%%%%%%
\beq\label{Eq:equality}
\rho_{\rm RH} =\rho_{\rm RH}^{\rm D} \,,
\eeq
where the left-hand side calculated only taking PBH evaporation as a source and the right-hand side for the inflaton decay. The radiation energy density at the end of PBH-driven reheating can be written as
\beq \label{Eq:pbhev}
\rho_{\rm RH}=\rho_R(\aev)\left(\frac{\aev}{\arh}\right)^4\simeq \rho_{\rm BH}(\aev)\left(\frac{\aev}{\arh}\right)^4 \,,
\eeq
and the inflaton energy density 
\bea \label{Eq:inflatonev}
\rho_\phi(\arh)=\rho_\phi(\ain)\left(\frac{\ain}{\arh}\right)^{3\,(1+w_\phi)}\,.
\eea
Now upon substitution of Eq.(\ref{Eq:bhenergyden}) into Eq.(\ref{Eq:pbhev}) and comparing with (\ref{Eq:inflatonev}), one can find
\bea
\left(\frac{\aev}{\arh}\right)^4=\beta^{\frac{4}{3\,w_\phi-1}}\left(\frac{\ain}{\aev}\right)^{\frac{12\,w_\phi}{1-3\,w_\phi}}\,.
\eea
Utilizing the above equation, $\rho_{\rm RH}$ can be written as,
\bea\label{Eq:rhorhpbh}
&&
\rho_{\rm RH}=48\pi^2\,\beta^{\frac{3\,(1+w_\phi)}{3\,w_\phi-1}}\left(\frac{\epsilon}{2\,(1+w_\phi)\,\pi\,\gamma^{3\,w_\phi}}\right)^\frac{2}{1-3\,w_\phi}
\nonumber
\\
&&
\left(\frac{M_P}{\Min}\right)^{\frac{6\,(1-w_\phi)}{1-3\,w_\phi}}\,M_P^4\,.
\label{Eq:pbhreheattemp}
\eea
Connecting Equations (\ref{Eq:rhorhpbh}), (\ref{Eq:equality}) and (\ref{Eq:rhorh1}), we have
\bea 
&&
y_\phi^c=\sqrt{\frac{8\pi}{\alpha_n}}\,\beta^{\frac{3\,(1-w_\phi)}{4\,(3w_\phi-1)}}\left(\frac{48\pi^2}{\lambda}\right)^{\frac{1-w_\phi}{4\,(1+w_\phi)}}
\\
&&
\left(\frac{\epsilon\,\gamma^{-3\,w_\phi}}{2\pi\,(1+w_\phi)}\right)^{\frac{1-w_\phi}{2\,(1-3\,w_\phi)\,(1+w_\phi)}}\left(\frac{M_P}{\Min}\right)^{\frac{3}{2}\frac{(1-w_\phi)^2}{(1-3w_\phi)\,(1+w_\phi)}}\,,
\nonumber
\eea
where $\alpha_n=\frac{2\,(1+w_\phi)}{(5-9\,w_\phi)}\sqrt{\frac{6\,(1+w_\phi)\,(1+3w_\phi)}{(1-w_\phi)^2}}$. The above equation is true for $n<7$. For $n>7$, similarly, instead of using Eq.(\ref{Eq:rhorh1}) employing Eq.(\ref{Eq:rhorh2}), one can find
\bea 
&&
y_\phi^c=\sqrt{-\frac{8\pi\,\beta}{\alpha_n}}\,(48\pi^2)^{\frac{3\,w_\phi-1}{6\,(1+w_\phi)}}\,\lambda^{\frac{w_\phi-1}{4\,(1+\,w_\phi)}}
\\
&&
\left(\frac{\epsilon\,\gamma^{-3\,w_\phi}}{2\pi\,(1+w_\phi)}\right)^{-\frac{1}{3\,(1+w_\phi)}}\left(\frac{M_P}{\Min}\right)^{\frac{w_\phi-1}{1+w_\phi}}\left(\frac{\rho_{\rm end}}{M_P^4}\right)^{\frac{5-9\,w_\phi}{12\,(1+\,w_\phi)}}\,.
\nonumber 
\eea

\vspace{0.2cm}

\section{Expression for $y_\phi^{\rm th}$}\label{appendixb}
The coupling strength $y_\phi^{\rm th}$ which ensures that the inflaton reheating happens before 
the evaporation process completes, 
can be determined by equating 
$\aev\sim \arh$, where expression 
for $\arh/a_{\rm end}$ is followed 
by Eq.(\ref{arhnl7}) for $n<7$ and 
Eq.(\ref{arhng7}) for $n>7$. And $\aev/a_{\rm end}$ followed by the expression (see, for instance, 
Ref.\cite{RiajulHaque:2023cqe})
\bea
\frac{\aev}{\ae}=
\left[\frac{(1+w_\phi)}{2 \sqrt{3}\,\epsilon}
\frac{\Min^3\sqrt{\rhoe}}{M_P^5}\right]^{\frac{2}{3(1+w)}}
\label{Eq:aev}
\eea
Now, comparing the above equations, for $n<7$, we have 
\bea\label{ytrnl7}
y_\phi^{\rm th}=\nu_1\left(\frac{M_P}{\Min}\right)^{\frac{3\,(1-w_\phi)}{2\,(1+w_\phi)}}\,,
\eea
and for $n>7$
\bea\label{ytrng7}
y_\phi^{\rm th}=\nu_2\left(\frac{\Min}{M_P}\right)^{\frac{1-3\,w_\phi}{1+w_\phi}}\left(\frac{\rho_{\rm end}}{M_P^4}\right)^{\frac{5-9\,w_\phi}{12\,(1+w_\phi)}}\,,
\eea
where $\nu_1=\sqrt{\frac{8\pi}{\alpha_n}}\left(\frac{1+w_\phi}{2\,\epsilon}\sqrt{\frac{\lambda}{3}}\right)^{\frac{w_\phi-1}{2\,(1+w_\phi)}}$ and $\nu_2=\sqrt{-\frac{8\pi}{\alpha_n}}\left(\frac{1+w_\phi}{2\sqrt{3}\,\epsilon}\right)^{\frac{1-3\,w_\phi}{3\,(1+w_\phi)}}\lambda^{\frac{w_\phi-1}{4\,(1+w_\phi)}}$.

\end{document}